\documentclass[usenames,dvipsnames]{article}
\usepackage{stmaryrd}
\usepackage{bm}
\usepackage[utf8]{inputenc}
\usepackage{bbm}
\usepackage{mathtools}
\usepackage{subcaption}
\usepackage[toc,page]{appendix}
\usepackage{amsmath,amsfonts,amssymb,amsthm,mathrsfs} 
\usepackage[a4paper,vmargin={2.4cm,2.8cm},hmargin={2.2cm,2.2cm}]{geometry} 
\usepackage[labelfont={bf}, margin=1cm]{caption} 
\usepackage{graphicx,graphics} 
\usepackage{epsfig} 
\usepackage{latexsym} 
\usepackage[english]{babel} 
\usepackage{enumerate}
\usepackage{changepage}
\usepackage{hyperref}

\newcommand{\Tr}[1]{\ensuremath{{\rm Tr}\,#1}}

\newcommand{\Diag}[1]{\ensuremath{{\rm Diag}#1}}
\newcommand{\sinc}[1]{\ensuremath{{\rm sinc}#1}}
\newcommand{\mat}[1]{\ensuremath{\mathbf{#1}}}
\newcommand{\vect}[1]{\ensuremath{\bm{#1}}}
\usepackage{geometry}
\usepackage{authblk}

\usepackage{authblk}
\title{Stability of large complex systems with heterogeneous relaxation dynamics}
\author[,1,2]{Pierre Mergny\thanks{\texttt{mergny.pierre@gmail.com}}}
\author[1]{Satya N. Majumdar \thanks{\texttt{satya.majumdar@universite-paris-saclay.fr}}}
\affil[1]{LPTMS,  CNRS,  Univ.   Paris-Sud,  Université  Paris-Saclay,  91405  Orsay,  France}
\affil[2]{Chair of Econophysics $\&$ Complex Systems, Ecole polytechnique, 91128 Palaiseau Cedex, France}
\date{}

\begin{document}

\maketitle

\begin{abstract}

We study the probability of stability of a large complex system of size 
$N$ within the framework of a generalized May model, which assumes 
a linear dynamics of each population size $n_i$ (with respect to its equilibrium
value): $ 
\frac{\mathrm{d}\, n_i}{\mathrm{d}t} = - a_i n_i - \sqrt{T} \sum_{j} 
J_{ij} n_j $. The $a_i>0$'s are the intrinsic decay rates,
$J_{ij}$ is a real symmetric $(N\times N)$ Gaussian random matrix and
$\sqrt{T}$ measures the strength of pairwise interaction between different 
species. Unlike in May's original homogeneous model, each species 
has now an intrinsic damping $a_i$ that may differ from one 
another. As the interaction strength $T$ increases, the system
undergoes a phase transition from a stable phase to an unstable phase
at a critical value $T=T_c$. We 
reinterpret the  probability of stability in terms of
the hitting time of the level $b=0$ of an associated 
Dyson Brownian Motion (DBM), 
starting at the initial position $a_i$ and evolving in `time' $T$.
In the large $N \to \infty$ limit, using this DBM picture, we are able 
to completely characterize $T_c$ for arbitrary density $\mu(a)$
of the $a_i$'s. For 
a specific flat configuration $a_i = 1 + \sigma \frac{i-1}{N}$, 
we obtain an explicit parametric solution for the 
limiting (as $N\to \infty$) spectral density for arbitrary $T$ and $\sigma$.
For finite but large $N$, we also compute the 
large deviation properties of 
the probability of stability  on the stable side $T < T_c$ using a Coulomb gas 
representation.

\end{abstract}

\section{Introduction}
\label{Sec:Introduction}

One of the main objectives in the study of large complex systems is to 
understand their stability properties. A major theoretical contribution 
to answer this hard question was made by Robert May in 1972 
\cite{may72}. Using a simple `toy' model May argued that large complex 
systems might become unstable as the system complexity (measured by the
strength of interactions between different units) increases. The 
seminal work of May was motivated by ecological questions at his time 
\cite{allesina2015stability}, but even today his results have found 
resonance among the study of large complex systems arising across
disciplines including economical 
sciences \cite{moran2019may}, neural networks 
\cite{sompolinsky1988chaos,wainrib2013topological}, gene 
regulations \cite{guo2020stability,PMID:27176315} to cite a few. May's approach will be 
discussed in detail below and consists in approximating the 
dynamics of the system by a set of linear coupled equations with random 
coefficients, and we refer to \cite{fyodorov2016nonlinear,biroli2018marginally,arous2020counting} for recent studies 
going beyond this linear approximation. 

In his original toy model, May considered a complex
system consisting of $N$ ecological species. To start with, each of the 
$N$ species is assumed to be in equilibrium with population $P_i^*$ ($i=1,2,\cdots, N$).
Consider first the case where the species are non-interacting and linearly stable. 
By linearly stable, one means that if the population size $P_i$'s are slightly perturbed
from their equilibrium values, then the deviation
$n_i(t)= P_i(t)-P_i^*$ for each $i$ evolves  in a deterministic manner as 
\begin{align}
\label{ls.1}
    \frac{\mathrm{d}  n_i(t)}{\mathrm{d}t} &= - n_i(t) \quad \text{for } i = 1, \dots,N \, . &&
\end{align}
For simplicity, May assumed identical intrinsic decay rate (set to be unity in Eq. (\ref{ls.1})) 
for each species, and this is what we call the \emph{homogeneous} relaxation hypothesis.  
Imagine now 
\emph{switching on} a pairwise interaction between the species, such that 
the dynamics is modified in the following way~\cite{may72}
\begin{align}
\label{eq:evolindex}
\frac{\mathrm{d}\, n_i(t)}{\mathrm{d}t} &= - n_i(t) - \sqrt{T} 
\sum_{j} J_{ij} n_j(t) \quad \text{for } i = 1, \dots,N \, \,  , && 
\end{align}
where $J_{ij}$ represents a pairwise interaction term 
which denotes the influence of the $j^{th}$ species on the 
relaxation dynamics of the $i^{th}$ species and 
$\sqrt{T}$ is a measure of the strength of this interaction.
The notation $\sqrt{T}$ for this interaction strength
may seem a bit strange at this stage, but we will see later
that $T$ will play the role of 'time' in the associated
Dyson Brownian motion picture. May's 
further assumption was to model this complex interaction matrix $J_{ij}$ as a random 
Gaussian matrix with real elements. The dynamics for $\vect{n}(t) = (n_1(t), \dots, 
n_N(t))$ in Eq. (\ref{eq:evolindex}) can be described in a compact matrix form as
\begin{align}
\label{may.1}
\frac{\mathrm{d}\vect{n}(t)}{\mathrm{d}t}  &= 
- \left(\mathbf{I} + \sqrt{T}\, \mathbf{J} \right)\, \vect{n}(t)  \, , &&
\end{align}
where $\mathbf{I}$ is the identity matrix and 
$\mathbf{J}$ is a real matrix with independent Gaussian entries.
To make further progress, May also assumed that the interaction matrix $J_{ij}$
is {\em symmetric}. In that case, the random matrix $J_{ij}$
coincides with the \emph{Gaussian Orthogonal Ensemble} (GOE) matrix
in the Random Matrix Theory (RMT) literature \cite{mehta2004random,forrester2010log}. Note that for a GOE matrix $\mat{J}$
has the same distribution as $-\mat{J}$, hence we have chosen an overall negative sign
in the interaction term in Eq. (\ref{eq:evolindex}) without any loss of generality. 

May’s equation (\ref{may.1}) then maps a dynamics question ``Is the multi-component system stable?”  
to a RMT question ``Are all the eigenvalues of the random matrix 
$\mathbf{B} =\mathbf{I} + \sqrt{T}\, \mathbf{J}$ positive?”. 
Using the properties of GOE matrices, May argued that strictly in the 
large $N$ limit (where all finite size fluctuations disappear), 
there exists a critical strength $T_c$ where the system 
undergoes a \emph{stability-instability phase transition} (sometimes known
as May-Wigner transition): for $T < T_c$ the 
system is stable while for $T>T_c$ it is always unstable. 
Using the well-known Wigner semi-circular law for the average eigenvalue
density of GOE eigenvalues as $N\to \infty$, May computed $T_c$
explicitly for this homogeneous model~\cite{may72}.
Thanks to the well-known properties of GOE matrices, one can go beyond
May's calculation of $T_c$ and even 
investigate the regime where $N$ is still large but finite and derive the 
behaviors of the typical and atypical fluctuations of the probability of stability 
 of the system~\cite{majumdar2014top}, as recalled briefly in the
next section.

One of the important ingredients in May's model (apart from the fact
that $J_{ij}$ is a GOE matrix) was to assume a homogeneous decay
rate for all species. In this paper, we address a simple question:
assuming that $J_{ij}$ is still a GOE matrix, how the May-Wigner
transition gets modified if one just makes the intrinsic decay rates
for the species \emph{heterogeneous}? This
is a natural and simple generalization of May's original toy model.
In this heterogeneous version, one just replaces
the identity matrix $\mathbf{I}$ in Eq. (\ref{may.1}) with an arbitrary diagonal matrix
with positive entries $\mathbf{A}=\mathrm{Diag}(a_1, \dots, a_N)$. Eq. (\ref{may.1})
now gets modified to 
\begin{align}
\label{may_hetero.1}
    \frac{\mathrm{d}\vect{n}(t)}{\mathrm{d}t}  &=-(\mathbf{A}+\sqrt{T}\, \mathbf{J})\vect{n}(t) =
- \mathbf{B}\, \vect{n}(t)\, ,  &&
\end{align}
where the effective relaxation matrix 
\begin{align}
\label{deform.1}
\mathbf{B}&= \mathbf{A} + \sqrt{T} \mathbf{J}= \sqrt{T}\left[\mathbf{J}+\frac{1}{\sqrt{T}}\, 
\mathbf{A}\right] \, , &&
\end{align}
can be interpreted as a deformation of the GOE matrix $\mathbf{J}$ by
an additive positive diagonal matrix $\mathbf{A}$, with $1/\sqrt{T}$ playing the
role of the strength of `perturbation'. As the `time' $T$ evolves, the
matrix $\mathbf{B}$ evolves from its `initial' value $\mathbf{A}$
and approaches a GOE matrix as $T\to \infty$.
In May's original homogeneous model where $\mathbf{A}=\mathbf{I}$, the matrix 
$\mathbf{B} $, for any strength parameter $\sqrt{T}$, is just a shifted GOE 
matrix. However, in the generic case $\mathbf{A} \neq \mathbf{I}$, the
spectrum of $\mathbf{B}$ is more complex and is a continuous interpolation between the 
spectrum of the matrix $\mathbf{A}$ and the spectrum of the rescaled 
GOE matrix, as a function of increasing $T$.
While deformed GUE (Gaussian unitary ensemble) models have been
studied extensively in the recent past with many applications
(see e.g.~\cite{krajenbrink2021tilted} and references therein),
here we obtain a natural example of a deformed GOE matrix.

For this heterogeneous May model, we expect again that in the limit $N \to 
\infty$, where there are no finite size fluctuations, there should a
critical value $T_c$ separating the stable ($T<T_c$) and the
unstable ($T>T_c$) phases. However, it turns out that the moment the intrinsic
diagonal positive rate matrix $\mathbf{A}$ differs from $\mathbf{I}$,
computing $T_c$ becomes highly nontrivial. In this paper we first develop
a general method to compute $T_c$ for arbitrary diagonal positive ${\mathbf{A}}$,
and then use it to calculate $T_c$ explicitly
for a particularly interesting case where the elements of $\mathbf{A}$
are distributed uniformly over a finite interval (we will refer to this
case as the \emph{flat initial
condition} since this corresponds to the value of $\mathbf{B}$ at ``time" $T=0$). 
This is the first main result of our paper. 

Next, for a general positive diagonal matrix $\mathbf{A}$, computing the average 
density profile of the eigenvalues of the deformed matrix $\mathbf{B}$, for arbitrary $T$, is also 
hard. However, for the `flat initial condition' described above, we
are able to compute analytically the average density of the eigenvalues of $\mathbf{B}$
in the large $N$ limit for arbitrary $T$ (in explicit parametric form), providing
our second main result.

Finally, for the same choice of $\mathbf{A}$ (the flat initial condition), we make the link 
with another ensemble, the deformed GUE, for which one can compute the joint density of 
eigenvalues for arbitrary $T$, going beyond just the average density. To the best of our knowledge, this 
provides a new RMT ensemble--a Coulomb gas in a harmonic potential, where the repulsive 
interaction between any pair of eigenvalues is a linear combination of logarithmic (as in 
the standard GUE) and log-sinh types. The RMT ensemble with only logarithmic (the standard 
GUE) or only log-sinh 
interaction~\cite{marino2005chern,marino2006matrix,dolivet2007chern,tierz2010schur,szabo2010chern,forrester2021global} 
have been studied before, but here we obtain naturally a linear combination of them as 
interaction. Such a mixed Coulomb gas is interesting to study in its own right. Moreover, 
this Coulomb gas approach also allows us to estimate, how for finite but large $N$, the 
probability of stability differs from $1$ on the stable side as one decreases $T$ below 
$T_c$ with $T_c-T\sim O(1)$ (we recall that strictly in the $N\to \infty$ limit, the 
probability of stability is exactly $1$ for $T<T_c$ by definition).

The rest of the paper is organized as follows: In Section 
\ref{Sec:MaysModel}, we recall in detail the derivation and properties 
of May’s original model. In particular, we describe in detail
the finite size effects on the May-Wigner transition, in terms of the Tracy-Widom 
distribution and the large deviation functions describing respectively 
the typical and atypical fluctuations of the system. We then describe 
the new main model with heterogeneous relaxation dynamics.  In Section 
\ref{sec:CriticalStrength_&_DBM}, we describe the main tools to perform 
the study of the heterogeneous model: the resolvent, and its link with 
Dyson Brownian Motion (DBM) and the Burgers' equation. This allows us to 
get the equation satisfied by the critical strength $T_c$ for a generic 
matrix $\mathbf{A}$.  In Section \ref{sec:parametricsol}, we show how 
one can get the parametric solution for the density, based again on the 
Burgers' equation. In Section \ref{sec:DeformedGUE_&_LDP}, we describe 
the deformed GUE with flat initial conditions and show how one can get 
the joint law density thanks to the Harish-Chandra-Itzykson-Zuber (HCIZ) 
integral. We then make the link with different models and show how one 
can get the large deviation function in the weakly stable phase for the 
original deformed GOE model with flat initial condition. Finally, we conclude in Section \ref{sec:Conclusion}. Some details of the computation are relegated in the Appendices.

\section{May's homogeneous model and its heterogeneous generalization}
\label{Sec:MaysModel}
\subsection{May's original homogeneous model}

The homogeneous May model has already been described in the introduction.
In this subsection, we show how $T_c$ for this model is computed in the strict $N\to \infty$
limit and then demonstrate how the probability of stability gets modified when $N$ is large, but finite.
Also, this recapitulation would be useful for understanding
the stability issues in the general setting of heterogeneous model that we will discuss in the
next subsection. 

\begin{figure}
     \centering
     \begin{subfigure}[b]{0.49\textwidth}
         \centering
         \includegraphics[width=\textwidth]{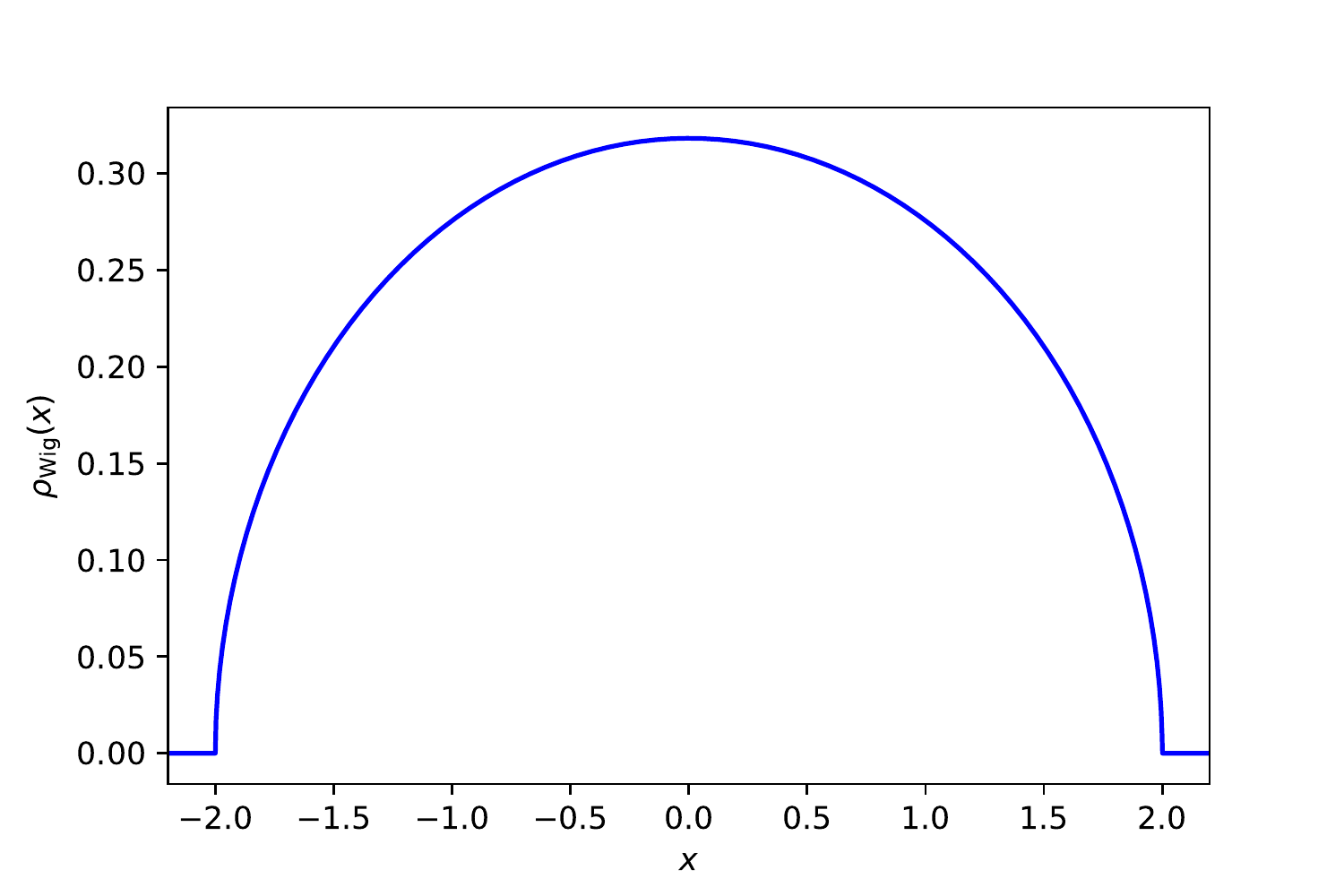}
         \label{fig:wig}
     \end{subfigure}
     \hfill
     \begin{subfigure}[b]{0.49\textwidth}
     \centering
         \includegraphics[width=\textwidth]{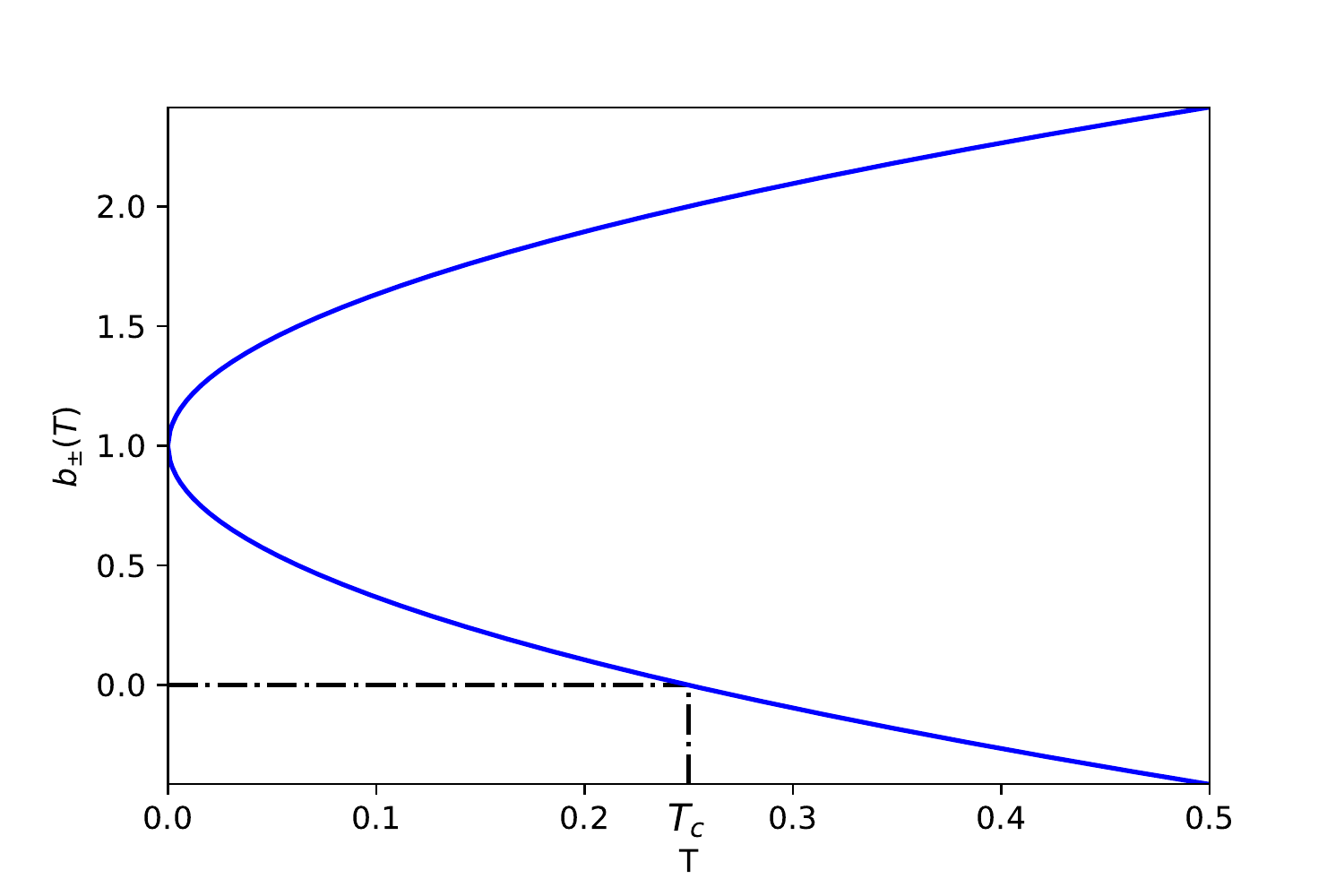}
         \label{fig:edgeMay}
     \end{subfigure}
    \caption{ \textbf{(Left)} Plot of the Wigner semi-circular 
distribution in Eq. (\ref{wigner_semi.1}). \textbf{(Right)} Plot of the 
two edges $b_{\pm}(T)=1\pm 2\, \sqrt{T}$  of the limiting density of 
the eigenvalues of the matrix $\mat{B}= I+ \sqrt{T}\, \mat{J}$ in Eq. (\ref{relax.1}), as a
function of $T$. The critical strength occurs at $T_c = \frac{1}{4}$, where the
lower edge $b_{-}(T)$ hits zero. }
    \label{fig:MayWigner}
\end{figure}

The deviations $n_i(t)=P_i(t)-P_i^*$ evolve via Eq. (\ref{may.1}) in the
original homogeneous model, where $\mat{J}$ is a GOE matrix.
Let us first briefly recall the properties of the GOE matrix.
The entries of a GOE matrix are symmetric, $J_{ij}=J_{ji}$
and the independent entries are distributed via
\begin{align}
\label{JL_GOE}
\mathcal{P}_N(\mat{J}) \mathrm{d}\mat{J} &\propto \mathrm{exp}\left[- \frac{N}{4} 
\Tr \mat{J}^2 \right] \mathrm{d}\mat{J} \, , && 
\end{align}
with $\mathrm{d}\mat{J} = \prod_{1 \leq i \leq j \leq N} \mathrm{d} J_{ij}$ being the Lebesgue 
measure on the space of symmetric matrices. The eigenvalues $\{\lambda_i\}$ of $\mat{J}$ are all real, and
it is well known \cite{wigner1951statistical} 
that the average density of eigenvalues,
\begin{align}
\label{density}
\rho(x,N)&= \frac{1}{N}\, \left\langle \sum_{i=1}^N \delta(\lambda_i-x)\right\rangle \, , &&
\end{align} 
converges in the $N \to \infty$ limit 
towards the \emph{Wigner semi-circular} distribution (see Fig. \ref{fig:MayWigner} (Left))
\begin{align}
\label{wigner_semi.1}
\rho(x, N\to \infty)\to \rho_{\mathrm{Wig}}(x) &= \frac{1}{2 \pi} \sqrt{4 - x^2} \quad 
\text{for} \quad  -2 \leq x\leq 2 \, . &&
\end{align}
As mentioned earlier, the joint density in Eq. (\ref{JL_GOE}) is manifestly invariant under the 
change $\mat{J} \to - \mat{J}$, so that  $\mat J \overset{\text{in law}}{=} - \mat J $.

The matrix form of May's equation (\ref{may.1}) reads
\begin{align}
 \frac{\mathrm{d}\vect{n}(t)}{\mathrm{d}t}  &=- \mathbf{B}\, \vect{n}(t) \, ,&&
\label{may.2}
\end{align}
where the effective relaxation matrix
\begin{align}
\mathbf{B}&= \mathbf{I}+\sqrt{T}\, \mathbf{J} \, , &&
\label{relax.1}
\end{align}
is just a shifted GOE. 
Let $\lambda_1<\lambda_2<\ldots <\lambda_N$ and 
$b_1(T)<b_2(T)<\ldots <b_N(T)$
denote the ordered eigenvalues of $\mat{J}$ and $\mathbf{B}$ in Eq. (\ref{relax.1}) 
respectively. Clearly,
\begin{align}
b_i(T)&= 1+ \sqrt{T}\, \lambda_i(T)\, , \,\, {\rm for}\, {\rm all}\, i=1,2,\ldots, N\, . &&
\label{bi.1}
\end{align}

\begin{figure}
     \centering
         \includegraphics[width= 0.55\textwidth]{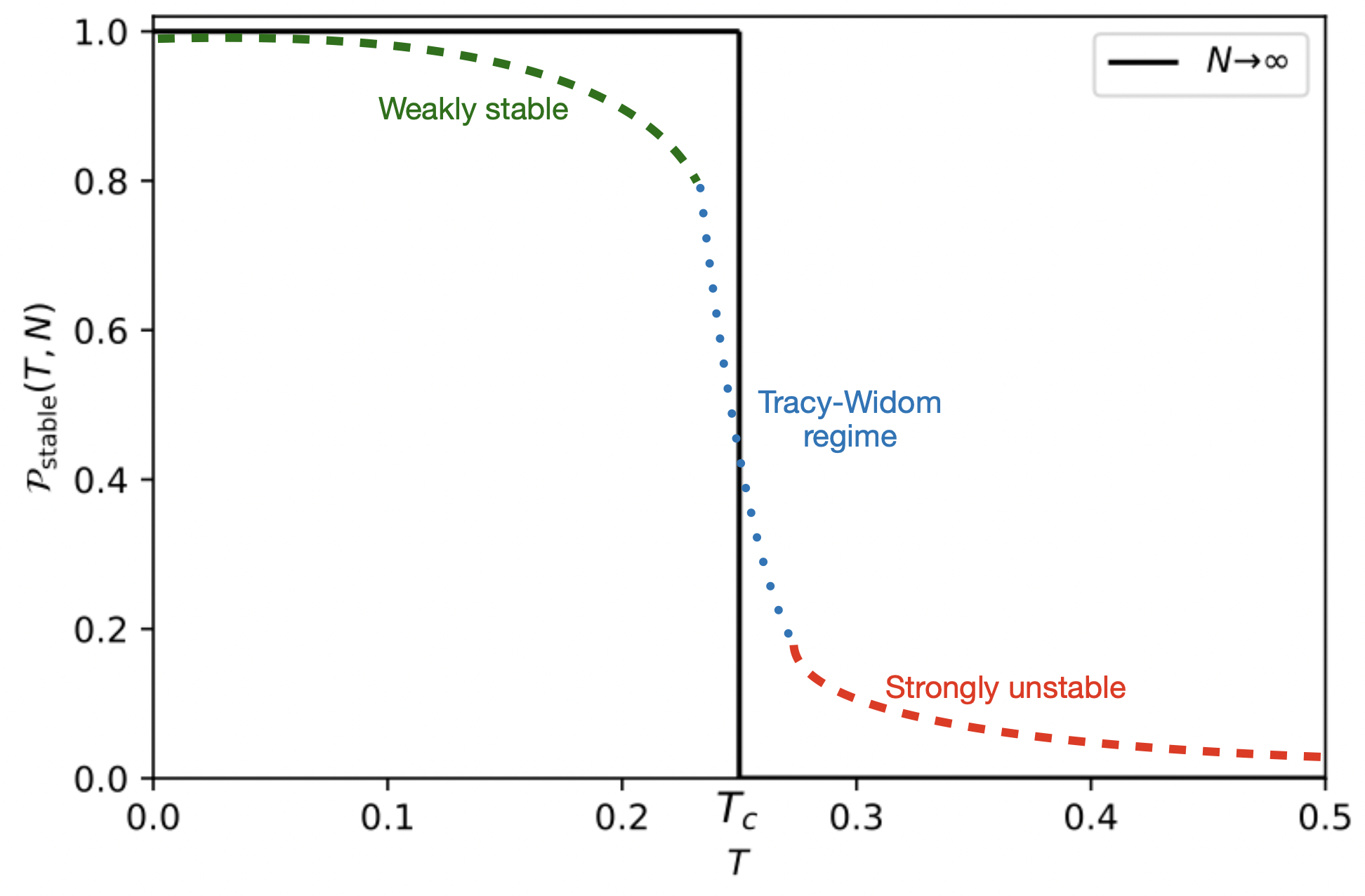}
    \caption{Sketch of the stability diagram in May's homogeneous model: the
 probability of stability $\mathcal{P}_{\mathrm{stable}}(T,N)$ as a function of $T$
for fixed large $N$.
The solid (black) line corresponds to the strictly $N \to \infty$ limit,
where $\mathcal{P}_{\mathrm{stable}}(T,N)$ is a step function with the step at $T_c=1/4$.
For finite but large $N$, this step function gets smoothened, as shown schematically
by the dashed and dotted lines. The dotted line around $T_c$, shows the
behavior of $\mathcal{P}_{\mathrm{stable}}(T,N)$ on a scale $|T-T_c|\sim O(N^{-2/3})$,
and has the Tracy-Widom form.
The dashed lines describe the behavior of $\mathcal{P}_{\mathrm{stable}}(T,N)$ 
when $|T-T_c|\sim O(1)$ and are described by the two large deviation behaviors on 
the two sides of $T_c$.} 
\label{fig:PvsT}
\end{figure}

One can now write down the condition for stability in terms of the ordered eigenvalues 
$\{b_i(T)\}$. From Eq. (\ref{may.2}), it is clear that the system is stable
if all eigenvalues of $\mathbf{B}$ are positive. Hence the probability of the
stability can be expressed, for fixed $T$ and $N$, as
\begin{align}
\label{eq:stab.0}
 \mathcal{P}_{\mathrm{stable}}(T,N)&=  
\mathrm{Prob} \left[ b_1(T) >0, \dots, b_N(T) >0 \right] \, ,&&
\end{align}
or equivalently since we have ordered the eigenvalues 
\begin{align}
\label{crit_stability}
\mathcal{P}_{\mathrm{stable}}(T,N)=  \mathrm{Prob}\left[b_1(T) >0 \right]
&= \mathrm{Prob}\left[\lambda_1> -\frac{1}{\sqrt{T}}\right]  \, , &&
\end{align}
where we used $b_1(T)=1+\sqrt{T}\, \lambda_1$ from Eq. (\ref{bi.1}).
For finite $N$, the value of $\lambda_1$, and hence that of $b_1(T)=1+\sqrt{T}\,\lambda_1$
fluctuates from sample to sample. However, strictly in the $N\to \infty$ limit,
we have seen before that the eigenvalues of $\mat{J}$ converge, almost surely, to
Wigner semi-circular law in Eq. (\ref{wigner_semi.1}). This means that, as $N\to \infty$,
all the eigenvalues $\{\lambda_i\}$ are supported within the finite interval $[-2,2]$.
Since $\lambda_1$ is
the lowest eigenvalue, it converges to the lower edge of the semi-circular, i.e.,
$\lambda_1\to -2$. Consequently, from Eq. (\ref{bi.1}), the eigenvalues $\{b_i(T)\}$ of $\mathbf{B}$
also converge to a shifted semi-circular law over the finite support
$[b_{-}(T), b_+(T)]$ (see Fig. \ref{fig:MayWigner} (Right)), where
\begin{align}
\label{b_pm}
b_{-}(T)&= 1-2\, \sqrt{T} \quad {\rm and} \quad b_+(T)= 1+2\,\sqrt{T}\, . &&
\end{align}
In particular, the lowest eigenvalue $b_1(T)$ converges to the lower edge as
$N\to \infty$, i.e., $b_1(T)\to b_{-}(T)=1- 2\,\sqrt{T}$.
This means that as $N\to \infty$, almost surely, $b_1(T)>0$ if $T<T_c=1/4$
and $b_1(T)<0$ if $T>T_c=1/4$. 
Thus, the probability of stability in Eq. (\ref{crit_stability}) also converges
to an $N$-independent form as $N\to \infty$
\begin{align}
\label{eq:stabNinf}
 \mathcal{P}_{\mathrm{stable}}(T,\infty) &= \left\{
    \begin{array}{ll}
        1 & \mbox{if } T < T_c = \frac{1}{4} \,  ,\\
        0  & \mbox{otherwise.}
    \end{array}
\right. &&
\end{align}
Thus, strictly in the $N\to \infty$ limit, the probability of stability, as a function of $T$,
approaches a `sharp' step function with the step located at
$T_c=1/4$, as shown in Fig. \ref{fig:PvsT}.

However, for finite but large $N$, this curve $\mathcal{P}_{\mathrm{stable}}(T,N)$
vs. $T$ will deviate from the step function (see Fig. \ref{fig:PvsT}). How
does the step function get modified for finite but large $N$?
To extract this information, we see from
Eq. (\ref {crit_stability}) that we need to know the probability distribution
of the lowest (minimum) eigenvalue $\lambda_1$ of an $(N\times N)$ GOE
matrix $\mat{J}$. Since for a Gaussian random matrix, the top eigenvalue
$\lambda_N$ has the same distribution as $-\lambda_1$ by symmetry, we can
equivalently express the probability of stability  in Eq. (\ref{crit_stability}) in terms
of the distribution of the top eigenvalue $\lambda_N$ of the GOE matrix, namely
\begin{align}
\label{crit_stability_top}
\mathcal{P}_{\mathrm{stable}}(T,N) &=  \mathrm{Prob}\left[b_1(T) >0 \right]
= \mathrm{Prob}\left[\lambda_1> -\frac{1}{\sqrt{T}}\right] =\mathrm{Prob}\left[\lambda_N< \frac{1}{\sqrt{T}}\right] \, . &&
\end{align}
Thus, we need to know how the top eigenvalue $\lambda_N$ of an $(N\times N)$ GOE matrix is
distributed for finite but large $N$.
At the time
of May's original work~\cite{may72}, this information was not available.
Currently, however, one knows a great deal about the distribution of
the top eigenvalue $\lambda_N$ of an $(N\times N)$ GOE matrix for finite but large $N$.
This information was used to estimate $\mathcal{P}_{\mathrm{stable}}(T,N)$
for finite but large $N$ in Ref.~\cite{majumdar2014top}, which we briefly recall below. 

\vskip 0.3cm

\noindent {\emph {Summary of the large $N$ behavior of the top eigenvalue $\lambda_N$ of an $(N\times N)$ GOE matrix.}} 
As mentioned earlier, the largest eigenvalue $\lambda_N$ converges to $2$ as $N\to \infty$, i.e.,
coincides with the right edge of the Wigner semi-circular density in Eq. (\ref{wigner_semi.1}). However,
for finite but large $N$, the random variable fluctuates around this right edge $2$. The scale of \emph{typical}
fluctuations is of order $O(N^{-2/3})$, where the probability 
distribution of $\lambda_N$, appropriately centered
and scaled, is described by the celebrated Tracy-Widom (GOE) distribution \cite{tracy2007nonintersecting,tracy1996orthogonal}. 
However, the Tracy-Widom form does not
describe the large fluctuations of $\lambda_N$ of $O(1)$ around $2$ and they are described by two
different large deviations form depending on whether $\lambda_N<2$ (left) \cite{dean2006large,dean2008extreme} or 
$\lambda_N>2$ (right) \cite{majumdar2009large}.
These three different regimes can be summarized by the following large $N$ behavior of the cumulative distribution
${\rm Prob}[\lambda_N<w]$ 
\begin{align}
\label{summary_top.1}
     \mathrm{Prob} \left[ \lambda_N <w \right]&\approx \left\{
    \begin{array}{lll}
        \mathrm{exp} \left[ - \frac{N^2}{2} \Psi_-(w) +o(N^2)  \right]  
& \mbox{for } w < 2  \mbox{ and } |w-2| \sim O(1)  \, ,\\
\\
        \mathcal{F}^{(1)}\left( N^{2/3}(w-2) \right) & 
\mbox{for } |w -2| \sim O(N^{-\frac{2}{3}}) \, , \\
\\
        1 -\mathrm{exp} \left[ - \frac{N}{2} \Psi_+(w) +o(N) \right] & 
\mbox{for } w>2 \mbox{ and } |w-2| \sim O(1) \, .
    \end{array}
\right. &&
 \end{align}
The Tracy-Widom (GOE) function $\mathcal{F}^{(1)}(x)$ can be expressed as
\begin{align}
\label{def:CDF_TW}
   \mathcal{F}^{(1)}(x) &= \mathrm{exp} \left[ - \frac{1}{2} \left( \int_x^{\infty} (s-x)
q^2(s) +q(s)\mathrm{d}s \right) \right] \, .&&
\end{align} 
where $q(s)$ is the Hasting-McLeod solution of the Painlevé II equation
\begin{align}
\label{eq:PainleveII}
    q''(s) &= 2q^3(s) + s q(s) \quad \text{such that} \quad q(s) \underset{s \to \infty}{\sim} \mathrm{Ai}(s)  \, .&&
\end{align}
The function $\mathcal{F}^{(1)}(x)$ has the leading order asymptotic behaviors
\begin{align}
\label{eq:TW_asymp}
     \mathcal{F}^{(1)}(x) &= \left\{
    \begin{array}{ll}
       \mathrm{e}^ {- \frac{1}{24} |x|^3 + o(|x|^{3})}   &, x \to - \infty \, , \\
\\
       1-\mathrm{e}^ { - \frac{2}{3} |x|^{3/2} +o(x^{3/2}) }   &, x \to  \infty  \, .
    \end{array}
\right.&&
\end{align}
The left and right large deviation functions, denoted respectively by $\Psi_{-}(w)$
and $\Psi_{+}(w)$ in Eq. (\ref{summary_top.1}), are also known explicitly
\cite{dean2006large,dean2008extreme,majumdar2009large} and read
 \begin{align}
     \Psi_-(w) &= \frac{1}{108} \left( 72 w^2 - 4w^4 - (15\sqrt{2}w + 2\sqrt{2}w^3)\sqrt{2w^2+6} +
27 \left(\ln 18 - 2 \ln \left( \sqrt{2}w + \sqrt{2w^2 +6}\right) \right) \right) \, , \,\, w\le 2\label{left.1}&& \\
          \Psi_+(w) &= \frac{w \sqrt{w^2 -4}}{2} - 2 \ln \frac{\sqrt{w^2 -4}+w}{2}\, ,  \, \,\, w\ge 2 .\label{right.1} &&
 \end{align}
The  large deviation functions have the following asymptotic behaviors near the edge $w=2$:
 \begin{align}
     \Psi_{-}(w) &\propto (2 - w)^3 \mbox{ for } w \to 2 \mbox{ and } w<2 \, , &&\\
     \Psi_{+}(w) &\propto (w - 2)^{\frac{3}{2}} \mbox{ for } w \to 2 \mbox{ and } w>2 \, . &&
\label{rate_asympt.1}
 \end{align}
One can check that these large deviation asymptotics as $w\to 2$ match smoothly with the Tracy-Widom tails
in Eq. \eqref{eq:TW_asymp}.

\vskip 0.3cm

\noindent {\emph {Large $N$ behavior of the probability of stability $\mathcal{P}_{\mathrm{stable}}(T,N)$ in the homogeneous May model.}}
Using the relation in Eq. (\ref{crit_stability_top}) one can then translate the large $N$ behavior of the cumulative
density function (CDF) of the
top eigenvalue ${\rm Prob}[\lambda_N<w]$ into the large $N$ behavior of $\mathcal{P}_{\mathrm{stable}}(T,N)$ in
May's homogeneous model. Setting $w=1/\sqrt{T}$ in Eq. (\ref{summary_top.1}), we see that the Wigner
edge $w=2$ corresponds to $T_c=1/4$ and the behaviors of the probability of stability
$\mathcal{P}_{\mathrm{stable}}(T,N)$ around $T_c=1/4$ for finite but large $N$ are described by 
\begin{align}
\label{summary_stable.1}
     \mathcal{P}_{\mathrm{stable}}(T,N) &\approx \left\{
    \begin{array}{lll}
        \mathrm{exp} \left[ - \frac{N^2}{2} \Phi_+\left(T \right) +o(N^2) \right]
& \mbox{for } T > T_c=1/4  \mbox{ and } |T-T_c| \sim O(1)  \, ,\\
\\
        \mathcal{F}^{(1)}\left( N^{2/3}\left(T^{-1/2}-2\right) \right) &
\mbox{for } |T -T_c| \sim O(N^{-\frac{2}{3}}) \, , \\
\\
        1 -\mathrm{exp} \left[ - \frac{N}{2} \Phi_-\left(T \right) +o(N)  \right] &
\mbox{for } T<T_c=1/4 \mbox{ and } |T-T_c| \sim O(1) \, ,
    \end{array}
\right. &&
 \end{align}
where $\mathcal{F}^{(1)}(x)$ is the Tracy-Widom (GOE) function. The rate functions $\Phi_{\pm}(T)$ are given by:
\begin{align}
\label{eq:PhiPsi}
  \Phi_{\pm}(T) &= \Psi_{\mp}\left(w =\frac{1}{\sqrt{T}}  \right) \, , && 
\end{align}
with $\Psi_{\mp}$ given by
Eqs. (\ref{left.1}) and (\ref{right.1}).
These behaviors are schematically sketched by the dashed-dotted lines in Fig. \ref{fig:PvsT} and describe
precisely how the sharp step function (for $N\to \infty$) gets modified for finite but large $N$.
In fact, the critical behavior around $T_c=1/4$ for finite $N$ in May's homogeneous model is similar to the so called `double scaling'
limit in various matrix models arising in lattice gauge theory and they all share a `third order' phase transition around
the critical point, as reviewed extensively in Ref. ~\cite{majumdar2014top}.

Let us remark that for finite but large $N$ and $T>T_c$, the 
 probability of stability $ \mathcal{P}_{\mathrm{stable}}(T,N)$ in Eq. (\ref{summary_stable.1}) in the large deviation 
regime $T-T_c\sim O(1)$
deviates only very slightly $\sim \exp[-O(N^2)]$ from its value $0$ when $N\to \infty$.
Thus, the $N\to \infty$ `unstable' phase remains `strongly' unstable when $N$ reduces from $\infty$. Hence we refer to
this phase as `strongly unstable' in Fig. \ref{fig:PvsT}. In contrast, for $T<T_c$, the deviation of 
$ \mathcal{P}_{\mathrm{stable}}(T,N)$ from its $N\to \infty$ value $1$ is of order $\sim \exp[-O(N)]$ which is much larger 
than the deviation $\sim \exp[-O(N^2)]$ on the other side, i.e., for $T>T_c$. Thus, for $T<T_c$,
the $N\to \infty$ `stable' phase, where the system was stable with probability $1$ when $N\to \infty$, 
is likely to change with a relatively higher probability when $N$ is reduced from $\infty$. 
Hence, in Fig. \ref{fig:PvsT}, we refer to the phase $T<T_c$ as the `weakly stable' phase.

\begin{figure}
     \centering
         \includegraphics[width= 1\textwidth]{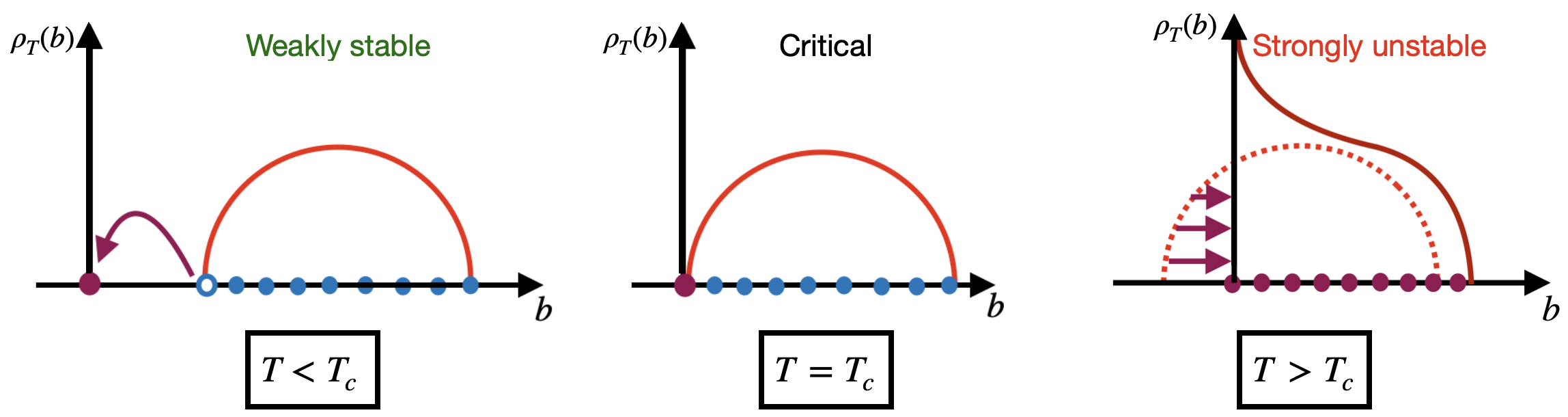}
         \label{fig:pulledpushed}
     \caption{Sketch of the two different processes leading to the different scaling in $N$ of the probability in Eq.  \eqref{summary_stable.1}. \textbf{(Left)}  For $T <T_c$, one needs to pull the lowest eigenvalue to the origin to make the system unstable, which does not change the equilibrium density. \textbf{(Center)} For $T=T_c$, the system is at the critical position where its lowest eigenvalue goes to $0$. \textbf{(Right)} For $T>T_c$, to make the system stable, one needs to push all eigenvalues above $0$. In this case, the gas rearranges itself and the equilibrium density is modified.   }
     \label{fig:PushPulled}
\end{figure}

Finally, we remark that these two different $N$ dependences of the large deviation behaviors
of $\mathcal{P}_{\mathrm{stable}}(T,N)$ on either side of $T_c$ admits a nice physical
interpretation in terms of the underlying log-gas picture of the eigenvalues of the relaxation
matrix $\mat{B}=I+ \sqrt{T}\, \mat{J}$, see Fig. \ref{fig:PushPulled}.
One can view the eigenvalues of the matrix $\mat{B}$ as a gas of $N$ particles living 
on the real line, confined by a harmonic potential and subject to a pairwise logarithmic repulsive interaction. For 
$T<T_c$, the system is asymptotically stable: this means all the eigenvalues $\{b_i(T)\}$ are
above $0$ for $T<T_c$ with probability $1$ in the $N\to \infty$ limit. To reduce this
probability from unity, i.e., to trigger an event that will make the system unstable for $T<T_c$,
one needs a rare configuration of charges for which  the lowest eigenvalue $b_1(T)<0$. This
can be achieved by \emph{pulling} the lowest eigenvalue $b_1(T)$ from its spectrum (whose lower edge is above $0$ for $T<T_c$)
to the value $0$. This costs energy of order $O(N)$ since one needs to disturb (pull) only one eigenvalue, without
disturbing the rest of the spectrum. Hence this explains the behavior $1- \mathcal{P}_{\mathrm{stable}}(T,N)\sim \exp[-O(N)]$
for $T<T_c$.
In contrast, for $T>T_c$, the system is 
asymptotically unstable, i.e., the lower edge of the spectrum of eigenvalues $\{b_i(T)\}$ is already
below $0$. To increase the stability, one needs to create a rare configuration where
one \emph{pushes} the whole gas of eigenvalues above $0$. Since this involves a re-arrangement
of $N$ particles in the Coulomb gas, it will cost energy of $O(N^2)$ (since each pair will contribute
when the whole gas is compressed from its equilibrium configuration). This explains the behavior
$\mathcal{P}_{\mathrm{stable}}(T,N)\sim \exp[-O(N^2)]$ for $T>T_c$. This `pulled' to `pushed'
phase transition occurs also in various lattice gauge models 
\cite{gross1980possible,wadia1980n,forrester2011non} where the `pulled' phase
corresponds to the `weak coupling' phase in gauge theory, while the `pushed' phase
corresponds to the `strong coupling' phase in gauge theory (for a review 
see \cite{majumdar2014top}).
Thus, the `stability-instability' phase transition in May's homogeneous model can also be
viewed as a `pulled-pushed' transition. The `stable' phase in May's model is the analogue
of the `weak coupling' phase of the gauge theory, while the `unstable' phase is the analogue
of the `strong coupling' phase of the gauge theory ~\cite{majumdar2014top}.

\subsection{heterogeneous relaxation dynamics}
\label{sec:model}

A natural extension of May's work is to drop the assumption that all the damping constants are equal and allow a spread in the 
distribution of the damping constants $a_i$'s, i.e., modify the evolution equation (\ref{eq:evolindex}) to 
\begin{align}
\label{eq:evolindex_het}
\frac{\mathrm{d}\, n_i(t)}{\mathrm{d}t} &= -a_i\, n_i(t) - \sqrt{T}
\sum_{j} J_{ij} n_j(t) \quad \text{for } i = 1, \dots,N \, \,  , &&
\end{align}
where the $a_i>0$'s and are not necessarily equal. In the matrix form, this can be written 
as in Eq. (\ref{may_hetero.1}) with
$\mat{A}$ being a diagonal matrix with positive entries $\{a_1, a_2, \ldots, a_N\}$. 
To keep the model simple, we will still assume that the matrix $\mat{J}$ in Eq. (\ref{may_hetero.1}) is a 
GOE matrix with density given in Eq. \eqref{JL_GOE}.  Since we will first study this generalized system in the $N \to \infty$ 
limit, we assume the empirical distribution of the $a_i$'s converges to a continuous distribution $\mu(a)$ whose support is included 
in the positive real axis (since we have assumed the $a_i >0$ to ensure stability without interactions). Thus,
$\mu(a)$ can be considered as the `initial' value of the deformed GOE matrix $\mat{B}$ at $T=0$. The homogeneous May model corresponds to the choice
of the `initial' condition
\begin{align}
\mu(a)&= \delta(a-1)\, . &&
\label{may_hom_init.1}
\end{align}  
Our main goal, in this paper, is to understand how the May-Wigner transition may get modified when there is a spread or
heterogeneity in the `initial' density $\mu(a)$.  

Starting from a given `initial' density $\mu(a)$ at $T=0$, the eigenvalues $\{b_i(T)\}$ of $\mat{B}$ will evolve
in `time' $T$. The first natural question is: for a general `initial' density $\mu(a)$, what is the limiting density  
$\rho_T(b)$ of the eigenvalues $\{b_i(T)\}$ at time $T$, in the $N\to \infty$ limit? For the special homogeneous
initial condition in Eq. (\ref{may_hom_init.1}), we have seen in the previous subsection that $\rho_T(b)$ is
a shifted semi-circular law with support over $b\in [1-2\, \sqrt{T}, 1+ 2 \sqrt{T}]$ at `time' $T$. For
a general $\mu(a)$, we will again expect that the limiting density $\rho_T(b)$ will have a finite support
$b\in [b_{-}(T), b_+(T)]$ at time $T$. If one can compute the location $b_{-}(T)$ of the lower edge
of the support of the limiting density as a function of $T$,   
then setting $b_{-}(T=T_c)=0$ will give us access to the
exact critical strength $T_c$ for an arbitrary `initial' condition $\mu(a)$. 

\begin{figure}
     \centering
         \includegraphics[width= 0.55\textwidth]{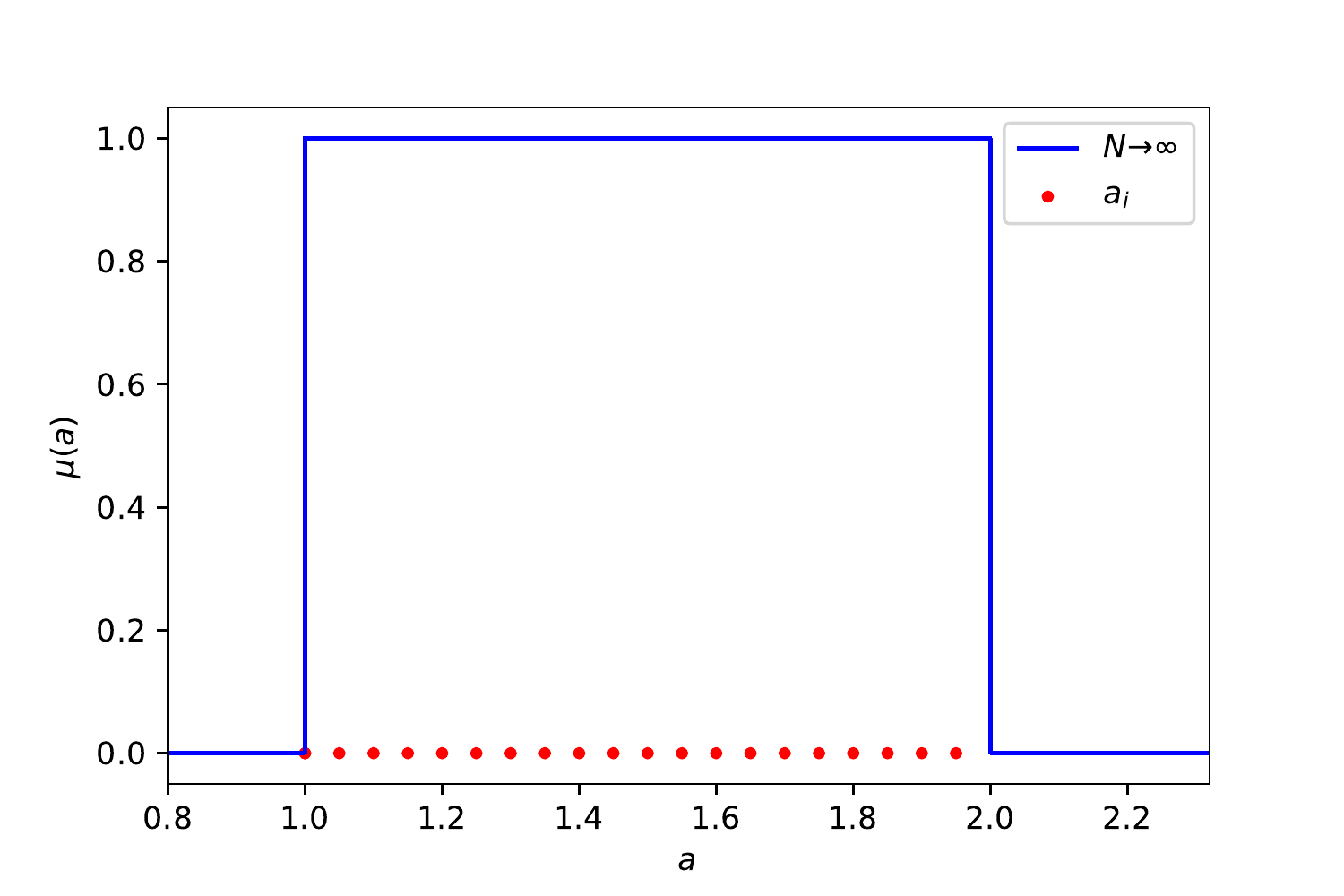}
         \label{fig:flatini}
     \caption{ Flat discrete initial configuration (in red) of the $a_i$ with $N=20$  in Eq. \eqref{flatIC} and their limiting flat density as $N \to \infty$ (in blue), for $\sigma=1$.}
     \label{fig:flatIC}
\end{figure}

Computing the limiting density $\rho_T(b)$ at $T$ for arbitrary `initial' density $\mu(a)$ seems rather hard.
However, one can make analytical progress for a specific choice of the `initial' values $b_i(T=0)=a_i$, 
\begin{align}
\label{flatIC}
a_i &= 1 + \sigma \frac{i-1}{N} \quad \text{for } i = 1,\dots,N \, ,&&
\end{align}
which we call the \emph{flat initial condition} since in the limit $N \to \infty$, the distribution of the $a_i$'s 
given by (\ref{flatIC}) converges towards the flat distribution $\mu(a)$ between $1$ and $1+\sigma$: 
\begin{align}
\label{flatIC2}
\mu(a) &=  \frac{1}{\sigma} \mathrm{\Pi}_{[1, 1+\sigma]}(a)  \, ,&&
\end{align}
where $\Pi_{[a,b]}(x)$ is the \emph{indicator function}:  $\Pi_{[a,b]}(x) = 1$ if $x$ is in $[a,b]$ and $0$ otherwise, 
see Fig. \ref{fig:flatIC}. The parameter $\sigma$ controls the width of this distribution and in particular the 
limit $\sigma \to 0$ corresponds to the homogeneous limit of May,  so that we can consider this new model as one parameter 
extension of  May's original model.

For this `flat initial condition', we are able to compute, in the $N\to \infty$ limit,
the limiting density $\rho_T(b)$ for all $T$. In particular, we will see in the next section that the 
precise knowledge of the lower edge $b_{-}(T)$ of its support
will enable us to compute the exact value of $T_c$ in this model. Furthermore, for finite but large $N$, we expect that for
the `flat initial condition', the  probability of stability $ \mathcal{P}_{\mathrm{stable}}(T,N)$ near its critical
point $T=T_c$ will have a qualitatively
similar behavior as in its homogeneous counterpart in Eq. (\ref{summary_stable.1}):
In particular, from
the general universality argument of the top eigenvalue of a GOE, we expect that the typical
fluctuation of $\mathcal{P}_{\mathrm{stable}}(T,N)$ will still be described by the Tracy-Widom (GOE) 
scaling function $\mathcal{F}^{(1)}(x)$ in the middle line of Eq. (\ref{summary_stable.1}). However,
the large deviation functions in the region $|T-T_c|\sim O(1)$, respectively in
the `unstable' and the `stable' side, are expected to be different in this heterogeneous `flat initial condition' model.
We will see in later sections that while we can compute
the rate function  on the `weakly stable' side, i.e., for $T<T_c$, computing
 the rate function in the `strongly unstable' phase remains a hard challenging problem even for the flat initial condition case.

\section{Critical strength and the hitting time of a Dyson Brownian Motion}
\label{sec:CriticalStrength_&_DBM}

The idea to characterize the critical strength in the general setting is 
to think of the parameter $T$ as a (fictitious) `time' variable of a 
well-known process called the Dyson Brownian Motion (DBM), described in 
Sec. \ref{Sec:DBM}. We can then obtain the time evolution of the associated 
resolvent, which satisfies the complex Burgers' equation, see Sec. 
\ref{sec:BurgersandTc}. Using the dynamics for the resolvent, we can derive in principle the critical strength $T_c$ for a general `initial' density $\mu(a)$, see Sec. \ref{sec:criticalT}. For the special case
of a `flat initial condition', we derive $T_c$ explicitly in
Sec. \ref{sec:criticalT}. Moreover, for this initial condition,
the resolvent can be solved explicitly giving access to the
full density $\rho_T(b)$, for arbitrary $T$, as discussed
in the next section (Sec. \ref{sec:parametricsol}). 

\subsection{Dyson Brownian Motion}
\label{Sec:DBM}

\begin{figure}
     \centering
     \begin{subfigure}[b]{0.49\textwidth}
         \centering
         \includegraphics[width=\textwidth]{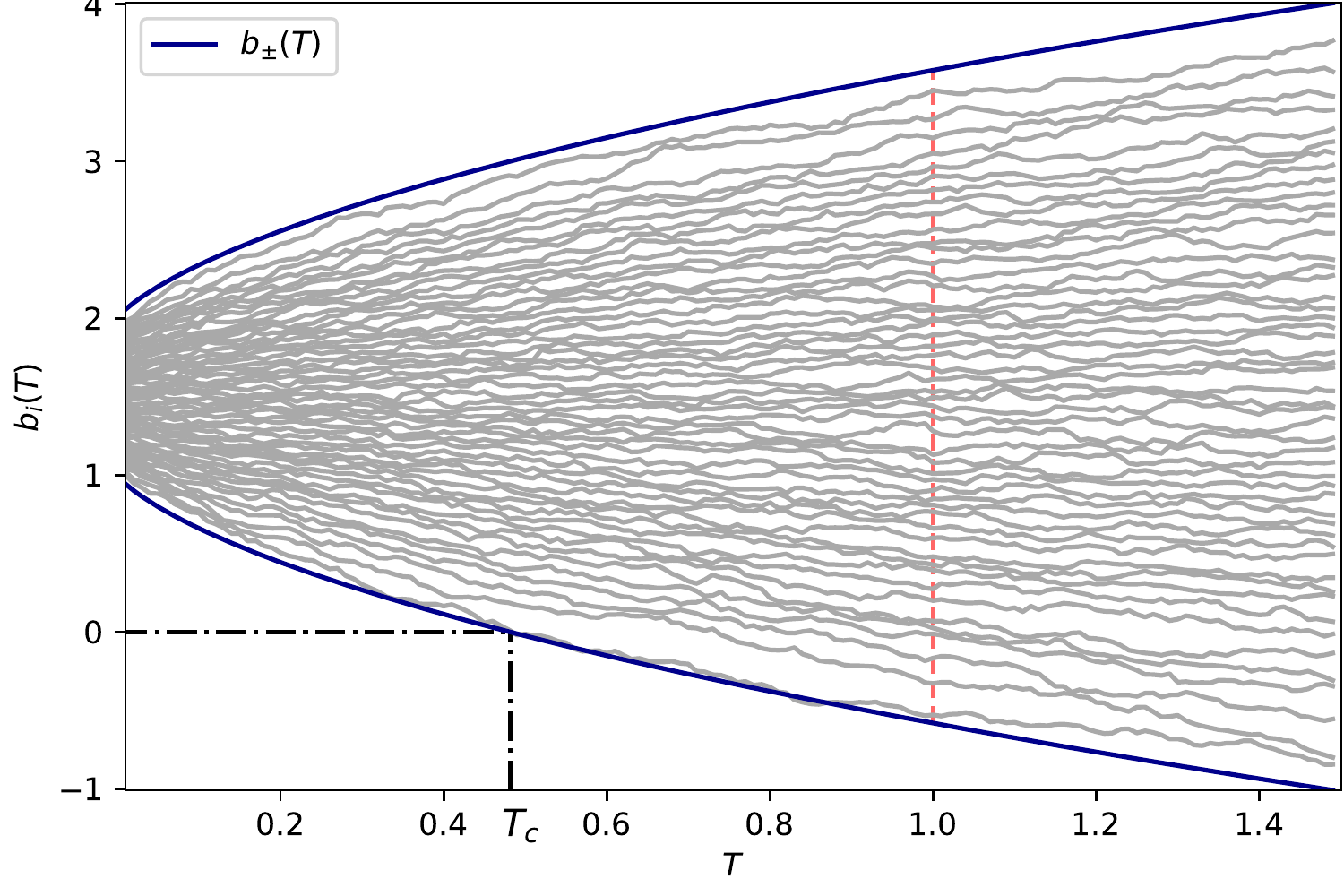}
         \label{fig:DBM}
     \end{subfigure}
     \hfill
     \begin{subfigure}[b]{0.49\textwidth}
         \centering
         \includegraphics[width=\textwidth]{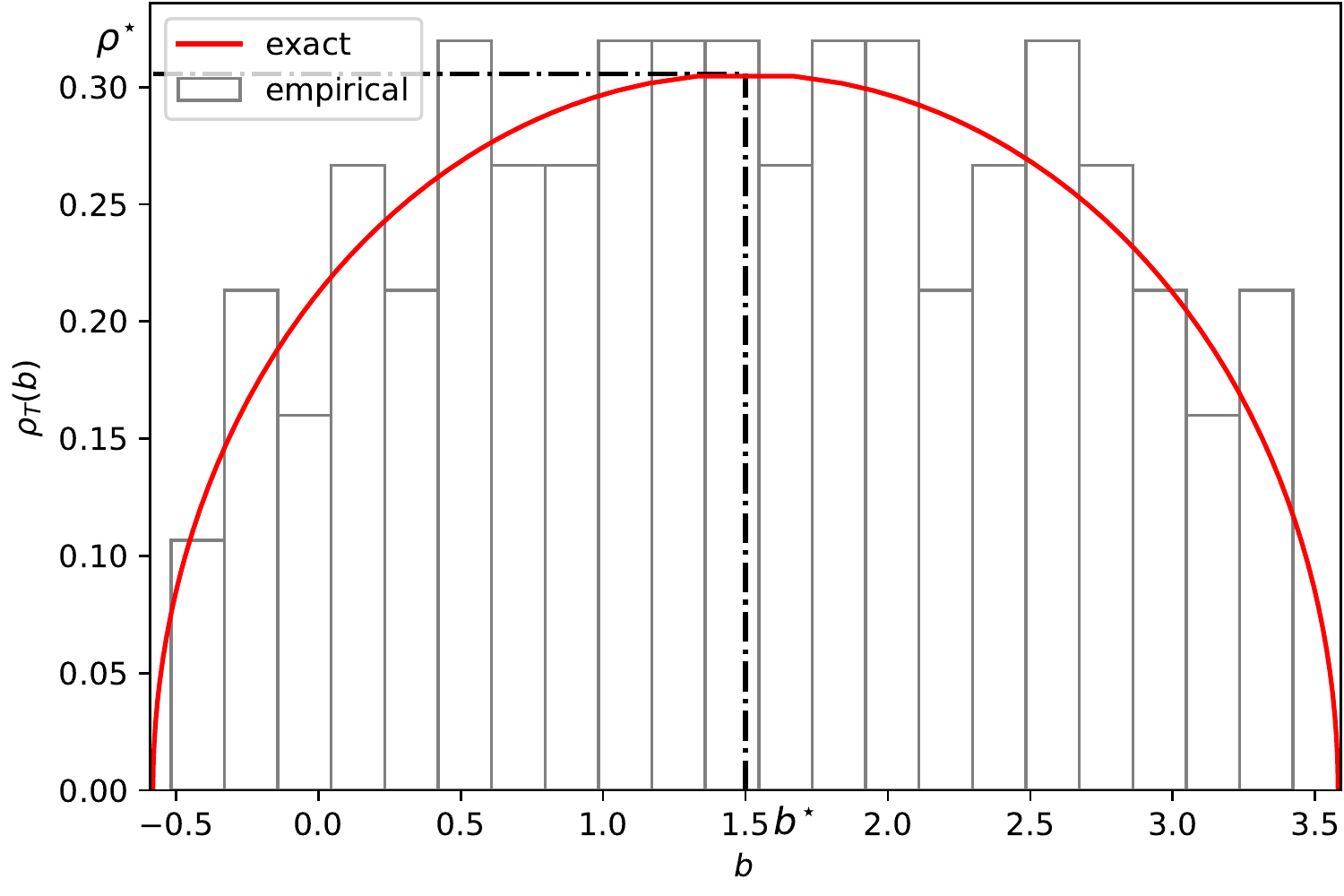}
         \label{fig:density}
     \end{subfigure}
     \caption{\textbf{(Left)} A representation of the DBM described by Eq. \eqref{DBM} at $\beta=1$ with the initial flat condition \eqref{flatIC} for $\sigma =1$ and $N=50$. In blue, the limiting curve for the bottom and top edges as $N \to \infty$. The dotted line corresponds to the value of $T=1$ of  the plot of the limiting density in the right. \textbf{(Right)} Plot of the limiting density  for the flat initial density using the parametric solution \eqref{eq:parametricsolution} for $\sigma =T =1$, compared to a histogram of the positions of the DBM at this time. }
     \label{fig:DBM_and_density}
\end{figure}

From the joint law of the elements of the GOE matrix \eqref{JL_GOE}, by 
doing the change of variable in Eq. \eqref{relax.1}, we get the joint law 
of the elements $B_{ij}$ of the matrix $\mat{B}$,
\begin{align}
\label{eq:prob_of_B}
    \mathcal{P}_N (\mat{B}) \mathrm{d}\mat{B} &\propto \mathrm{exp} 
\left\{ - N\, \Tr \left[ \frac{(\mat{B} - \mat{A})^2}{4T} \right] 
\right\} \mathrm{d}\mat{B} \, , &&
\end{align}
with $\mathrm{d}\mat{B}= \prod_{1 \leq i \leq j \leq N} \mathrm{d}B_{ij}$. Since $\mat{A}$ is diagonal, this can be equivalently written as: 
\begin{align}
\label{eq:prob_of_B.2}
 \mathcal{P}_N (\mat{B}) \mathrm{d}\mat{B} &\propto \prod_{i=1}^N \mathrm{exp} \left[ - \frac{N}{4T} (B_{ii} - a_i)^2 \right] \mathrm{d}B_{ii} \prod_{j:j>i} \mathrm{exp} \left[ - \frac{N}{2T} B_{ij}^2 \right] \mathrm{d}B_{ij}\, .&&
\end{align}
Under this form, one recognizes the propagator of the Brownian motion for a particle starting at time $T=0$ at the position $x_0$ and evaluated at time $T$:
\begin{align}
\label{eq:GaussPropgator}
    \mathcal{P} \left( x, T | x_0, 0 \right) &=  \mathrm{exp} \left[ - \frac{1}{4DT} ( x - x_0)^2 \right] \, , &&
\end{align}
with the diffusion constant $D=\frac{1}{N}$. Under this framework, we 
can naturally interpret the strength parameter $T$ as a time variable: 
each element of the matrix $\mat{B}$ evolves according to a Brownian 
motion until time $T$, starting at $a_i$ for the diagonal element 
$B_{ii}$ and starting at $0$ for the off-diagonal elements. 
At each instant $T$, one can diagonalize the matrix $\mat{B}$
and obtain its eigenvalues $\{b_i(T)\}$ which are real.
It is then natural to ask how these eigenvalues $\{b_i(T)\}$
evolve with time $T$.
Using a second order perturbation theory as in quantum mechanics, 
Dyson \cite{dyson1962brownian} showed that they follow what is now known 
as the ($\beta =1$) \emph{Dyson Brownian Motion} (DBM), namely
\begin{align}
\label{DBM}
\frac{\mathrm{d} b_i(T)}{\mathrm{d}T } &= \frac{1}{N} \sum_{j:j \neq i} 
\frac{1}{b_i(T) - b_j(T) } + \sqrt{2 D}\, \eta_i(T) \, , &&
\end{align} 
starting from the initial condition, 
\begin{align}
\label{eq:IC_DBM}
b_i(0) &= a_i \, . &&
\end{align}
In Eq. \eqref{DBM}, $\eta_i(T)$, for each $i$, is
an independent Gaussian white noise with zero mean and 
correlator $ \langle\eta_i(t) \eta_j(t') \rangle = \delta_{ij} \, 
\delta(t-t')$. The trajectory of a $\beta=1$ DBM, evolving with $T$, is 
represented in Fig. \ref{fig:DBM_and_density} (Left). 
As Dyson did, one can redo the computation for Hermitian and 
symplectic matrices $\mat{J}$. For each case 
(symmetric/Hermitian/symplectic), one gets again the 
dynamics \eqref{DBM}, with a diffusion constant $D=\frac{1}{N \beta}$, 
where $\beta =1,2,4$ is the Dyson index characterizing
these three standard ensembles of Gaussian matrices. 
In particular the case $\beta =2$ will be useful later which 
also corresponds to the dynamics of $N$ \emph{vicious walkers}, 
or non-intersecting Brownian motions, see \cite{forrester2011non,karlin1959coincidence,grabiner1999brownian,rambeau2011dis,grela2021non}. For large $N$,  
one can therefore interpret the critical value $T_c$ as the hitting time 
of the barrier at $b=0$ of such a DBM, when the lower edge $b_{-}(T)$
hits the level $0$, see Fig. \ref{fig:DBM_and_density} (Left). 
One then needs first to compute the lower edge $b_-(T)$ of the 
DBM in the large $N$ limit and then compute the critical 
strength (or time) by setting 
\begin{align}
\label{carac_HT_Tc}
b_-(T_c)& =0 \, .&&
\end{align}

\subsection{The resolvent and the complex Burgers' equation}
\label{sec:BurgersandTc}

The main tool to perform the computation of the edges of an evolving
density of eigenvalues $\{b_i(T)\}$ via Eq. \eqref{DBM} is to introduce 
\emph{the resolvent}, a key transform in RMT. The derivation of the properties of the resolvent needed in this section are recalled in the Appendices. For $z \in 
\mathbb{C}$ and $z \neq b_i(T)$ the resolvent is defined by

\begin{align}
\label{def:finiteresolvent}
g_N(z,T) &= \frac{1}{N} \Tr (z\, \mat{I}-\mat{B})^{-1}=
\frac{1}{N} \Tr \left( z\,\mat{I} - \mat{A} - 
\sqrt{T} \mat{J} \right)^{-1} = 
\frac{1}{N} \sum_{i=1}^N \frac{1}{z - b_i(T)} \, . &&
\end{align}
In the large $N \to \infty$ limit, the sum is replaced by an integral,
defined for all $z$ in the complex plane outside the support 
$[b_{-}(T), b_+(T)]$ of the density on the real axis,
\begin{align}
\label{def:resolvent}
g(z,T) &= g_{\infty}(z,T) = 
\int_{b_{-}(T)}^{b_{+}(T)}  
\frac{\rho_T(b)}{z -b} \mathrm{d}b  \, , &&
\end{align}
with $\rho_T(b)$ denoting the limiting density of the DBM at time $T$. 
From the knowledge of a resolvent, one gets the corresponding density, 
using the Socochi-Plemelj formula (see Appendix \ref{sec:Ap:def_res}),
\begin{align}
\label{prop:Plemelj}
\rho_T(b) &= \frac{1}{\pi} \mathfrak{Im} \,  g(b - \mathrm{i} 0^+,T) \, ,&&
\end{align}
where $\mathfrak{Im}$ denotes the imaginary part.
The lower and the upper edges $b_{\mp}(T)$ 
of the density $\rho_T(b)$
can be extracted from the resolvent $g(z,T)$ by applying
the following general prescriptions (see Appendix \ref{Sec:App:edge_and_res} for
the derivation).

\vskip 0.3cm

\begin{itemize}

\item First, define the inverse function $z(\mathrm{g})$ of the resolved 
$g(z)$, i.e.,
$z(g(z))=z$. We have suppressed the $T$ dependence of $g(z,T)$
for convenience. For example, for the semi-circular density in 
Eq. (\ref{wigner_semi.1}), the resolvent $g(z)= (z\pm \sqrt{z^2-4})/2$
is well known. Its inverse function is then $z(\mathrm{g})=
\mathrm{g}+1/\mathrm{g}$.

\item Next, find the roots of $z'(\mathrm{g})=0$ 
where $z'(\mathrm{g})=dz(\mathrm{g})/d\mathrm{g}$.
In general, this equation will have multiple roots. For a 
density confined in a single interval on the real line, 
this has typically two roots, denoted by $g_*$ (the lower one) and $g^*$ (the upper one).  
For example, for the semi-circular distribution, 
$z'(\mathrm{g})=1-1/\mathrm{g}^2=0$ 
gives two roots
$\mathrm{g}_*=-1$ and $\mathrm{g}^*=1$. 
The smallest root is $\mathrm{g}_*=-1$ and the largest one is $\mathrm{g}^*=1$.

\item The lower edge $b_{-}(T)$ of the support of the density is then given by
\begin{align}
\label{prop:edge_from_resolvent}
b_{-}(T) &= z(\mathrm{g}_*) \, , &&
\end{align}
Similarly, the upper edge of the support is given by the other root, i.e.,
$b_{+}(T)= z(\mathrm{g}^*)$.
For example, for the semi-circular law, one gets $b_{-}=-2$ and $b_{+}=2$
which indeed are respectively the
lower and the upper edge of the support $[-2,2]$ of 
the density in Eq. (\ref{wigner_semi.1}).

\end{itemize}

\begin{figure}
     \centering
         \includegraphics[width= 0.65\textwidth]{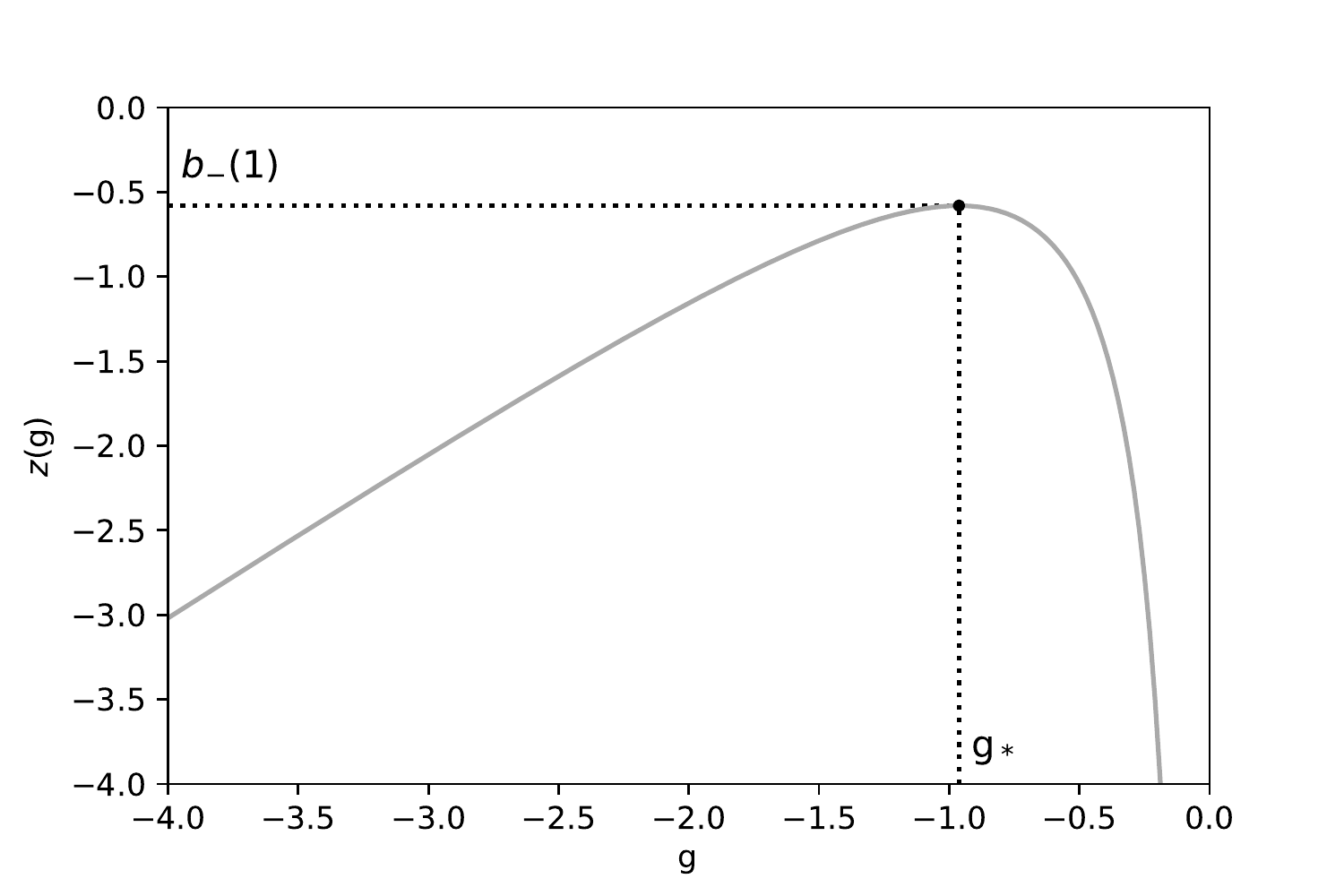}
     \caption{Representation of the inverse function $z(\mathrm{g})$ 
of the resolvent $g(z,T)$ for $T= \sigma = 1$.  
If one starts from the origin $g=0$ and goes to the left, 
the function $z(\mathrm{g})$ increases until it reaches the point 
$g_*$ from which one can get the lower edge $b_{-}(1) 
\approx -0.580457$.}
     \label{fig:inverse_resolvent}
\end{figure} 

Next, one needs to find the equation of motion describing the resolvent 
from which we can get the lower edge by  Eq. \eqref{prop:edge_from_resolvent}. 
Since the resolvent in Eq. \eqref{def:finiteresolvent} is a functional of the positions of
particles evolving according to the DBM \eqref{DBM}, the idea is to apply Ito's lemma in Eq. \eqref{DBM} to get the 
stochastic evolution for the resolvent. Taking the limit $N \to \infty$ 
the noise term vanishes and one can show that $g(z,T)$ is the solution 
of the \emph{complex inviscid Burgers' equation} 
\cite{menon2012lesser,blaizot2010universal},
\begin{align}
\label{def:BurgersEq}
\partial_T g(z,T) + g(z,T) \partial_z g(z,T) &=0 \, ,&&
\end{align}
evolving from the initial condition $g(z,0) = g_0(z) = 
\int \mathrm{d}a \frac{\mu(a)}{z -a}$. Using the method of 
characteristics, see for example Sec. 3 of \cite{grela2021non},  
the solution can be expressed in a parametric form as 
\begin{align}
\label{eq:Burgerscarac}
g(z,T) &= g_0 (\xi) \, ,&&
\end{align}
with
\begin{align}
\label{def:xi}
\xi &= z- Tg_0(\xi) \, .&&
\end{align}
For $z$ and $T$ fixed and given $g_0$, 
one first needs to solve Eq. \eqref{def:xi} for $\xi$ 
and then inject the solution in Eq. \eqref{eq:Burgerscarac}. 
Conversely, from Eq. \eqref{def:xi} with $\xi$ fixed, 
one can express $z$ as an implicit function of $\xi$
\begin{align}
\label{def:zofxi}
    z(\xi) &= \xi + T g_0(\xi) \, .&&
\end{align}
The idea would be to eliminate $\xi$ from Eqs. (\ref{eq:Burgerscarac}) 
and (\ref{def:zofxi}) to obtain $g(z,T)$ as a function of $z$ for fixed $T$.
However, in practice, this is not always easy, as we will see shortly.

\subsection{Critical strength} 
\label{sec:criticalT} 

We now have all the necessary ingredients to compute  
the critical 
strength $T_c$. Since the lower edge is given by  Eq. 
\eqref{prop:edge_from_resolvent}, we need to first solve
$z'(\mathrm{g})=0$  and find its lowest root, see Fig. \ref{fig:inverse_resolvent} for an illustration.
The equation $z'(\mathrm{g})=0$ is equivalent to
\begin{align}
\label{dzdg.2}
    \frac{\mathrm{d}z(\xi)}{\mathrm{d}\xi} 
\frac{\mathrm{d}\xi}{\mathrm{d}g} &= 0\, . && 
\end{align}
In general the term $\frac{\mathrm{d}\xi}{\mathrm{d}g}$ is non-zero, 
hence this is equivalent to solve 
\begin{align}
\label{dzdxi}
    \frac{\mathrm{d}z(\xi)}{\mathrm{d}\xi} &= 0 \, . &&
\end{align}
Using the expression \eqref{def:zofxi} for $z(\xi)$, one gets
\begin{align}
\label{prop:xi_star}
1 +  T g'_{0} \left(  \xi_{*}(T) \right)& = 0 \, ,&&
\end{align}
where $\xi_{*}(T)$ denotes the lowest root of Eq. (\ref{prop:xi_star}).
Injecting this $\xi_{*}(T)$ back into Eq. (\ref{def:zofxi}) and using Eq.
\eqref{prop:edge_from_resolvent} gives the lower edge
\begin{align}
\label{prop:bottomedge}
b_{-}(T) &=  \xi_{*}(T) + T \,  g_{0} \left( \xi_{*}(T) \right) \, , &&
\end{align}
where  $\xi_{*}(T)$ is obtained from Eq. (\ref{prop:xi_star}).
Finally, setting $b_{-}(T_c)=0$ gives $T_c$.

To summarize, we have again the following large $N \to \infty$ behavior
of the probability of stability with arbitrary initial density $\mu(a)$, 
\begin{align}
\label{prop:ProbaVST}
 \mathcal{P}_{\mathrm{stable}}(T,\infty) &= \left\{
    \begin{array}{ll}
        1 & \mbox{if } T < T_c \, ,\\
        0  & \mbox{otherwise.}
    \end{array}
\right.&&
\end{align}
where now the critical strength $T_c$, which
implicitly depends on $\mu(a)$,  
is obtained from the solution of the transcendental equation 
\begin{align}
\label{prop:equationTc}
 \xi_{*}(T_c) + T_c \, g_{0} \left( \xi_{*}(T_c) \right)& = 0  \, ,&&
\end{align}
with $\xi_{*}(T_c)$ is given in Eq. \eqref{prop:xi_star}. 
Our algorithm for determining $T_c$, for arbitrary initial
density $\mu(a)$, thus follows three principal steps:

\vskip 0.3cm

\begin{itemize}

\item Given $\mu(a)$, we first determine the initial resolvent
$g_0(z)=\int \mathrm{d}a \frac{\mu(a)}{z -a}$.

\item Once we have $g_0(z)$,
we solve Eq. (\ref{prop:xi_star}) and determine $\xi_*(T)$.

\item Next we inject this $\xi_*(T)$ in the  
transcendental equation \eqref{prop:equationTc} and solve it to determine $T_c$.

\end{itemize}

For example, in May's original homogeneous model, we have
$\mu(a)=\delta(a-1)$. This gives, $g_0(z)=1/(z-1)$.
Substituting this in Eq. (\ref{prop:xi_star}), we get two roots, and
the lowest root gives $\xi_{*}(T)=1-\sqrt{T}$. Substituting this
in Eq. \eqref{prop:equationTc} gives $1-2 \sqrt{T_c}=0$, and hence
$T_c=1/4$. Our method, outlined above, holds for arbitrary $\mu(a)$
and in the next subsection, we show
that for the flat initial condition with $\mu(a)$ given in Eq. (\ref{flatIC2}),
the general procedure described above can be carried
out explicitly, thus providing a nontrivial generalization
of May's homogeneous initial condition. 

\vskip 0.3cm

\begin{figure}
     \centering
         \includegraphics[width= 0.65\textwidth]{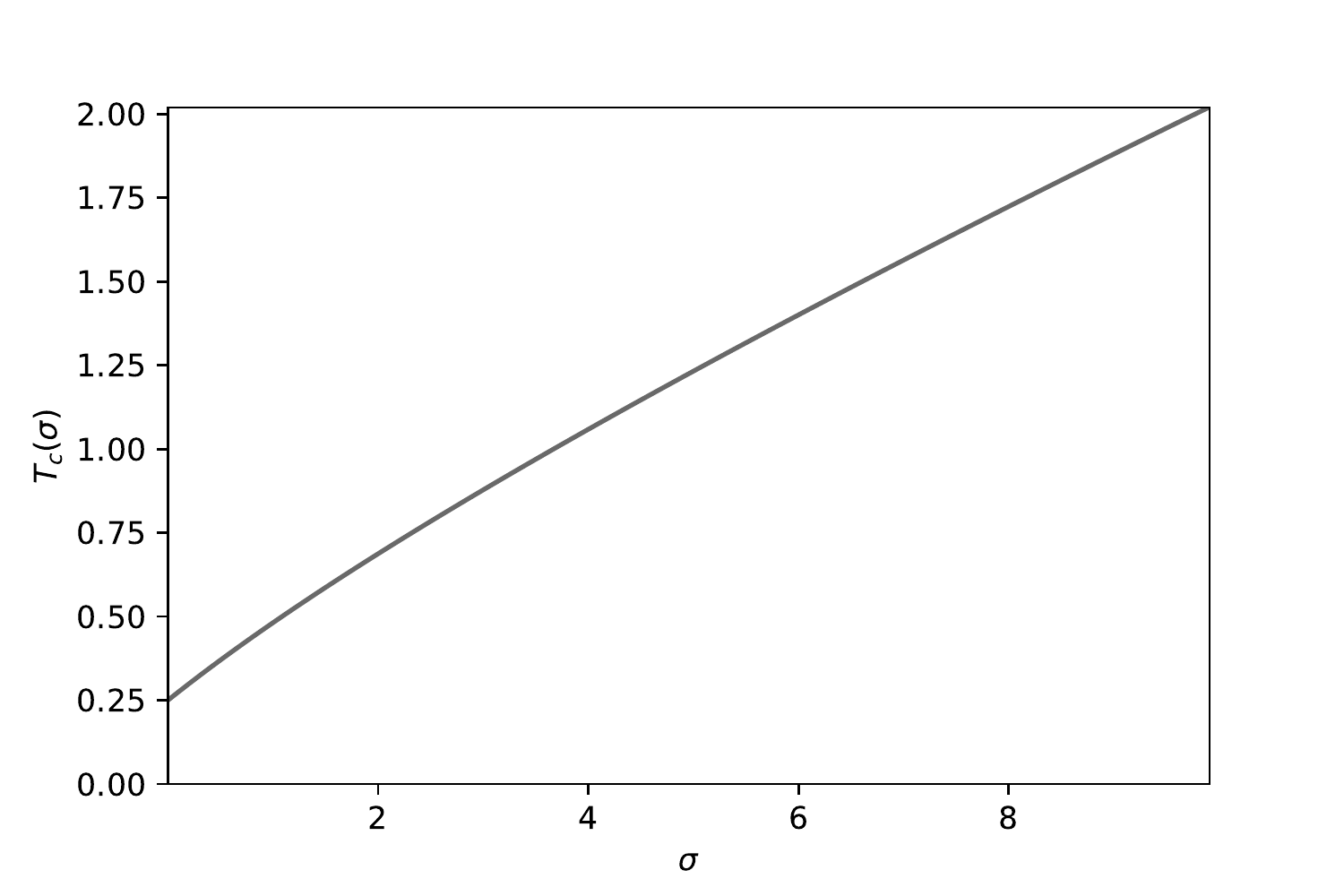}
         \label{fig:Tc}
     \caption{Plot of the critical strength $T_c$ given in Eq. \eqref{Tc_flat.1},  as a 
function of the spread $\sigma$.}
     \label{fig:T_versus_sigma}
\end{figure}

\noindent{\emph {Critical strength for the flat initial condition:}} 
As a nontrivial example, we now consider the flat initial condition
with $\mu(a)$ given in Eq. \eqref{flatIC2}. In this case, the initial 
resolvent is given by: 
\begin{align}
\label{eq:resolventg0}
g_0(z) &= \frac{1}{\sigma} \int_{1}^{1+\sigma} \frac{\mathrm{d}a}{z - a} 
= \frac{1}{\sigma} \ln \left( \frac{z-1}{z-1-\sigma} \right) \, ,&&
\end{align}
and its derivative is given by
\begin{align}
\label{eq:dg0dz}
g_0'(z) &= - \frac{1}{(z-1)(z-1-\sigma)} \, .&&
\end{align}
Using Eq. \eqref{prop:xi_star}, $\xi_*(T)$ satisfies the quadratic equation 
\begin{align}
\label{eq:quadraticXi}
\left( \xi_*(T) -1) \right) \left( \xi_*(T) -1 - \sigma \right) &=T \, , &&
\end{align}
whose lowest solution is given by
\begin{align}
\label{eq:Xi}
 \xi_*(T) &=  1 + \frac{\sigma}{2} - \frac{\sigma}{2} 
\sqrt{1+ \frac{4 T}{\sigma^2}} \, .&&
\end{align}
Using Eq. \eqref{prop:bottomedge}, the lower edge at fixed $T$ is given by
\begin{align}
\label{bl_flat.1}
b_{-}(T) &=   1 + \frac{\sigma}{2} - 
\frac{\sigma}{2} \sqrt{1+ \frac{4 T}{\sigma^2}} +
\frac{T}{\sigma} \ln \frac{ \sqrt{1+ 
\frac{4 T}{\sigma^2}}-1}{1 + \sqrt{1+ \frac{4 T}{\sigma^2}}}\, .&&
\end{align}
Setting $b_{-}(T_c)=0$ in Eq. (\ref{bl_flat.1}) gives $T_c$. However,
it is not easy to solve explicitly this transcendental equation.
To proceed further, we first write Eq. (\ref{bl_flat.1}) in a more 
compact form, 
\begin{align}
\label{bm_flat.1}
b_{-}(T) &= 1 - \sigma h \left( \frac{4T}{\sigma^2 } \right) \, ,&&
\end{align}
where the scaling function $h(u)$ is given by
\begin{align}
\label{hu_flat.1}
h(u) &=  \frac{\sqrt{1+u} -1 }{2}  + 
\frac{u}{2}\, \ln \left( \frac{1+\sqrt{u+1}}{\sqrt{u}}  \right) \, . &&
\end{align}
This function admits the following asymptotic behaviors near the origin and at infinity: 
\begin{align}
\label{prop:Assymptforh}
h(u)&\sim \left\{
    \begin{array}{ll}
       \frac{1}{4}\left( 1 + 2 \ln 2 - \ln u \right) u & \mbox{for } 
u \to  0 \, ,\\
\\
     \sqrt{u}- \frac{1}{2}  + \frac{1}{6} u^{-\frac{1}{2}}&  
      \mbox{for } u \to  \infty \, .
    \end{array}
\right.&&
\end{align}
Setting $b_{-}(T_c)=0$ in Eq. (\ref{bm_flat.1}) gives
\begin{align}
\label{hu.1}
h\left(\frac{4 T_c}{\sigma^2}\right) &= \frac{1}{\sigma}\, .&&
\end{align}
Hence we can write
\begin{align}
\label{Tc_flat.1}
T_c (\sigma) & = \frac{\sigma^2}{4} u\left( \frac{1}{\sigma} \right) \, ,&&
\end{align}
where $u(h)$ is the inverse function of $h(u)$. Since the function $h(u)$
is explicit in Eq. (\ref{hu_flat.1}), its inverse function $u(h)$
can be easily plotted. Thus, we can plot $T_c$ in Eq. (\ref{Tc_flat.1})
as a function of the spread $\sigma$,
as shown in Fig. \ref{fig:T_versus_sigma}. 
The asymptotic behaviors of $T_c$ for small and large $\sigma$
can also be derived using Eq. (\ref{prop:Assymptforh}) and are given by 
\begin{align}
\label{prop:AssymptforT}
T_c(\sigma)&\sim \left\{
    \begin{array}{ll}
     \frac{1+\sigma}{4}  & \mbox{for } \sigma \to  0 \, ,\\
\\
      \frac{\sigma}{\ln \left( \frac{\sigma}{4} \right)}&  \mbox{for } \sigma \to  \infty \, .
    \end{array}
\right.&&
\end{align}
In particular, we recover as expected the limit 
$T_c = \frac{1}{4}$ of May's original model 
for $\sigma \to 0$.  In the limit $\sigma \to \infty $, we find 
$T_c \to \infty$ from Eq. (\ref{prop:AssymptforT}), which
indicates that for large $\sigma$, the system is always stable, 
regardless of the value of the strength parameter $T$. This is an
interesting result which perhaps
could not have been guessed a priori.

\section{Parametric solution for the density with flat initial condition}
 \label{sec:parametricsol}

The goal of this section is to obtain an expression for the limiting 
density $\rho_T(b)$ of the eigenvalues of the matrix $\mat{B}$ for the 
flat initial condition \eqref{flatIC}, at arbitrary time $T$. 
The idea is to rely again on the 
complex Burgers' equation \eqref{def:BurgersEq} for the resolvent.  As 
we will see, the density $\rho_T(b)$ cannot be easily expressed in terms 
of known analytical functions. However, it can be expressed in
an easily plottable parametric form.

We start with the two basic equations satisfied by the resolvent $g(z,T)$,
namely the solution of the complex Burger's equation
in Eq. (\ref{eq:Burgerscarac}) and Eq. (\ref{def:zofxi}). For easy reading,
let us re-write these two equations together
\begin{align}
g(z,T) &= g_0 (\xi) \label{gg0} && \\
z(\xi) &= \xi + T\, g_0(\xi) \, . \label{zg0} &&
\end{align}
The idea is to eliminate the auxiliary variable $\xi$ between
these two equations and express $g$ as a function of $z$, for a fixed $T$.

To proceed, we start with the initial resolvent
\begin{align}
g_0(\xi) &= \int \frac{\, \mu(a)}{\xi-a}\mathrm{d} a\, . &&
\label{g0_init.1}
\end{align}
Suppose we could invert this equation and write $\xi$ as a function of $g_0$
\begin{align}
\xi &= z_0(g_0(\xi))\, . &&
\label{init_inv.1}
\end{align}
Thus $z_0(.)$ is just the inverse function of $g_0(\xi)$ in 
Eq. (\ref{g0_init.1}). Substituting Eq. \eqref{gg0} in Eq. \eqref{init_inv.1} gives
\begin{align}
\xi&= z_0(g(z,T))\, . &&
\label{xi_init.1} 
\end{align}
Using this relation in Eq. (\ref{zg0}) and further using $g_0 (\xi)=g(z,T)$,
Eq. (\ref{zg0}) reduces to 
\begin{align}
z &= T\, g(z,T)+ z_0(g(z,T))\, . &&
\label{eq:zofg}
\end{align}
Thus, for fixed $T$, if we know the initial inverse function $z_0(.)$, we have,
in principle, a closed equation for $g(z,T)$. 
From the expression (\ref{eq:resolventg0}) of the initial resolvent $g_0$ in
the flat initial condition case, 
its inverse function $z_0(g)$ is given by: 
\begin{align}
\label{eq:z0}
z_0(g) &= 1+ \sigma + \frac{\sigma}{\mathrm{e}^{\sigma g} -1} \, , &&
\end{align}
Substituting this in Eq. (\ref{eq:zofg}), we then have a closed equation for
the resolvent $g(z,T)$ at any time $T$
\begin{align}
z&= T\, g(z,T)+ 1+ \sigma + \frac{\sigma}{\mathrm{e}^{\sigma g(z,T)} -1}\, . &&
\label{resolvent_T.1}
\end{align}

Solving explicitly $g(z,T)$ from this transcendental equation does
not seem feasible, unfortunately.
To derive the density $\rho_T(b)$ from this resolvent $g(z,T)$
using Eq. (\ref{prop:Plemelj}), we set
$z = b - \mathrm{i} 0^+$, with $b$ between the two edges 
$b_{\pm}(T)$. Then by Eq. \eqref{prop:Plemelj} 
we have $g(b - \mathrm{i}0^+,T) =  u + \mathrm{i} \pi \rho $,
where $u$ is the real part of the resolvent. 
For simplicity, we have used the shorthand notation
$u \equiv  u(b,T)$ and $\rho \equiv \rho_T(b)$. 
Identifying the real and the imaginary parts of (\ref{resolvent_T.1}),
we get a pair of coupled equations 
\begin{align}
\label{eq:RealandIm1}
 \left\{
    \begin{array}{ll}
      b &= 1 + \sigma + Tu  + \sigma \, \mathfrak{Re} \left[  \dfrac{ 1}{\mathrm{e}^{ \sigma ( u + \mathrm{i} \pi \rho )}-1}  \right] \, , \\
      0 &= T\pi \rho+ \sigma \,  \mathfrak{Im} \left[\dfrac{ 1}{\mathrm{e}^{ \sigma ( u + \mathrm{i} \pi \rho )}-1}   \right] \, .\\
    \end{array}
\right. &&
\end{align}
One can multiply the numerator and denominator inside the brackets by 
$\mathrm{e}^{ \sigma ( u - \mathrm{i} \pi \rho )}-1$, to get the real 
and imaginary parts of the function inside the brackets, and the system 
can then be written as
\begin{align}
\label{eq:RealandIm2}
 \left\{
    \begin{array}{ll}
        b &=  1+ \sigma + T u  + 
\sigma  \dfrac{ \cos (\sigma \pi \rho) 
\mathrm{e}^{\sigma u} -1 }{\mathrm{e}^{ 2\sigma u} - 
2 \cos(\pi \sigma \rho)\mathrm{e}^{ \sigma u} +1} \, , \\
      0 &= T \pi \rho  - \sigma  \dfrac{ \sin(\pi \sigma \rho)  
\mathrm{e}^{ \sigma u}  }{\mathrm{e}^{ 2\sigma u} 
- 2 \cos( \pi \sigma \rho)\mathrm{e}^{\sigma u} +1} \, . \\
    \end{array}
\right. &&
\end{align}

Ideally, the goal would be to eliminate $u$ from these pair of equations
and express $\rho\equiv \rho_T(b)$ as a function of $b$, for fixed $T$.
Let us first consider the simple case of May's homogeneous model, i.e.,
the limit $\sigma\to 0$. In this limit, Eq. (\ref{eq:RealandIm2}) reduces to
\begin{align}
\label{eq:s0}
 \left\{
    \begin{array}{ll}
        b &=  1+ T u  + \frac{u}{u^2+\pi^2\, \rho^2}\, , \\
        \\
      0 &= T \pi \rho  - \frac{\pi\,  \rho}{u^2+\pi^2\, \rho^2} \, . \\
    \end{array}
\right. &&
\end{align}
Eliminating $u$ from these pair of equations, one immediately gets
the shifted Wigner semi-circular density
\begin{align}
\label{shifted_Wigner.1}
\rho_T(b){\Large {|}}_{\sigma=0} &= \frac{1}{2\pi T}\, \sqrt{4 T- (b-1)^2}\, , &&
\end{align}
supported over the interval $b\in \left[1-2\,\sqrt{T}, 1+2\, 
\sqrt{T}]\right]$. Thus, in May's homogeneous model, starting
from the initial condition $\mu(a)=\delta(a-1)$, the density
of eigenvalues $b_i(T)$'s, at any time $T>0$, is of the
shifted Wigner semi-circular form in Eq. (\ref{shifted_Wigner.1}). 

\begin{figure}
     \centering
     \begin{subfigure}[b]{0.49\textwidth}
         \centering
         \includegraphics[width=\textwidth]{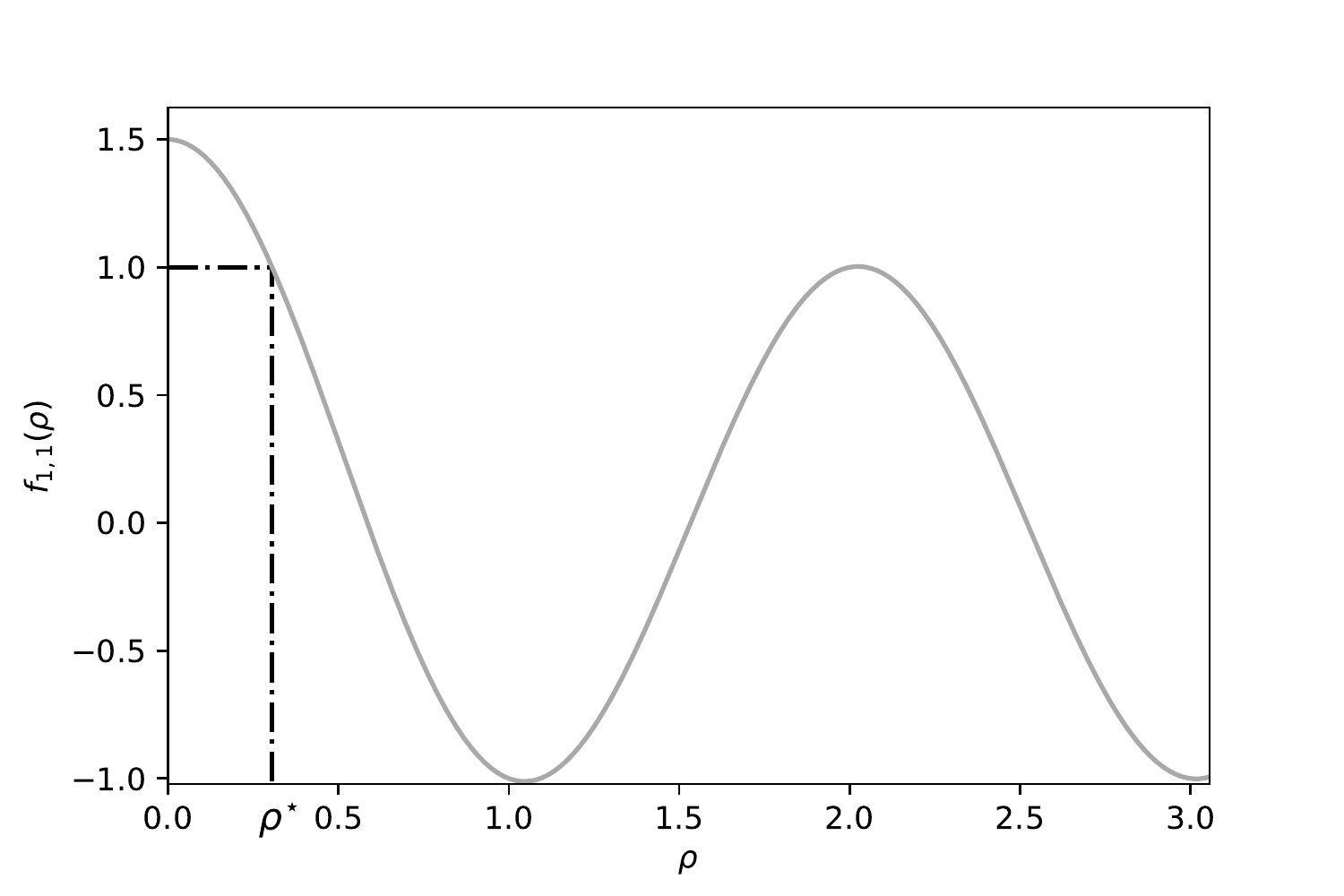}
         \label{fig:fofrho}
     \end{subfigure}
     \hfill
     \begin{subfigure}[b]{0.49\textwidth}
         \centering
         \includegraphics[width=\textwidth]{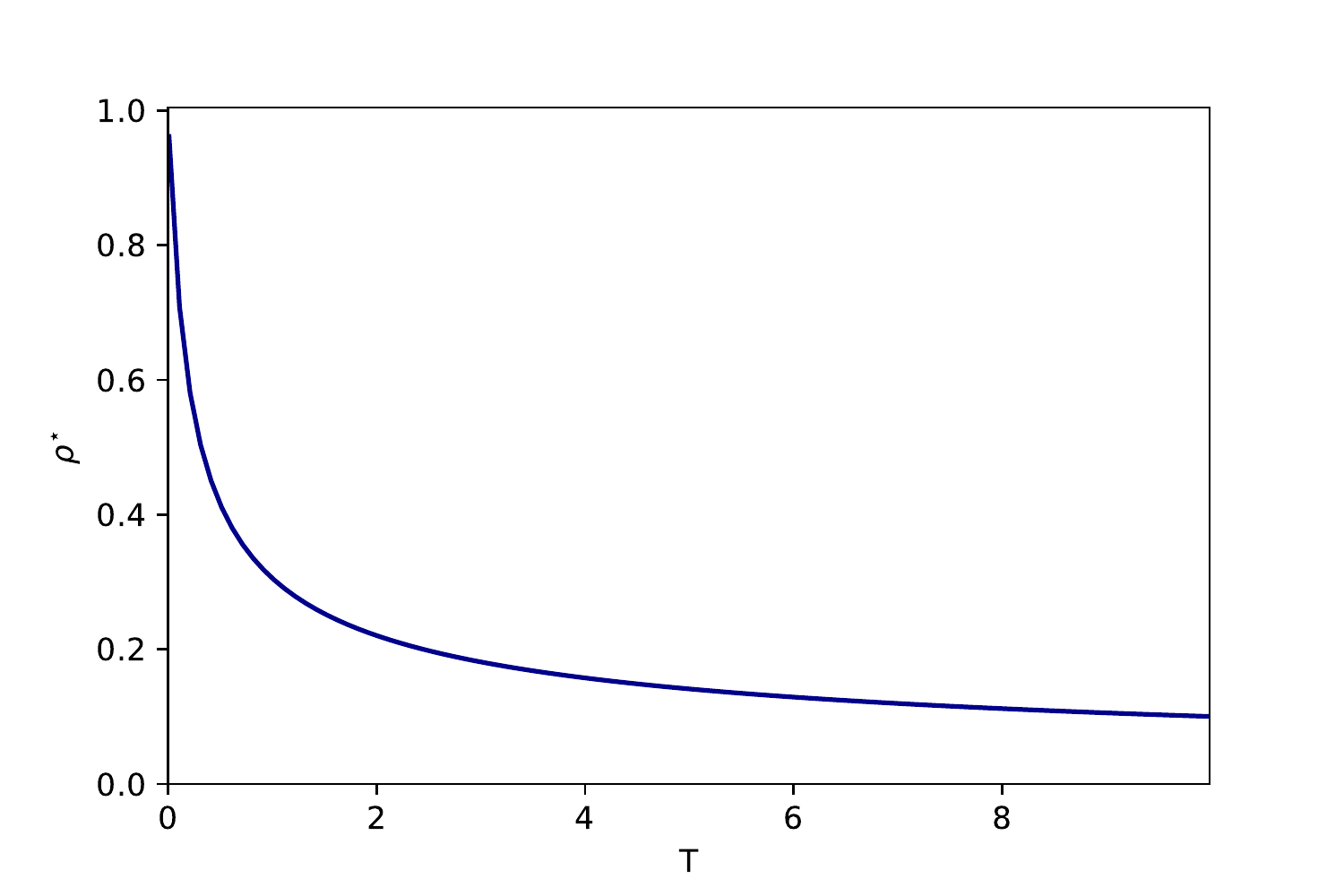}
         \label{fig:rhomax}
     \end{subfigure}
     \caption{ \textbf{(Left)} Plot of $f_{\sigma,T}(\rho)$  given in Eq. \eqref{def:f}, as a function of $\rho$ for $\sigma = T =1$. Only the part on the left of the value $\rho^{\star} \approx 0.305637 $ contribute to the parametric solution of the density. \textbf{(Right)} Plot of the maximum $\rho^{\star}$ described by Eq. \eqref{eq:rhomax.2} as a function of $T$, for $\sigma=1$.}
     \label{fig:f_and_rhomax}
\end{figure}

For general $\sigma>0$, eliminating $u$ from Eq. (\ref{eq:RealandIm2})
and expressing $\rho_T(b)$ explicitly (as in the $\sigma=0$ case)
seems difficult. Instead, for a general $\sigma>0$, one
can obtain the solution parametrically as follows.
We note that
the top equation of (\ref{eq:RealandIm2}) is a parametric 
expression for $b(u,\rho)$. 
The idea is to eliminate the dependency on $u$ by working a bit on
the bottom equation of (\ref{eq:RealandIm2}). 
To do so, let us denote by $w = \mathrm{e}^{\sigma u}$, and
then from the bottom equation of (\ref{eq:RealandIm2}) $w$ satisfies a 
quadratic equation, 
\begin{align}
\label{eq:SecondOrderw}
\frac{w^2}{2} - w\left( \frac{\sigma^2}{2T} \mathrm{\sinc}(\pi \sigma \rho)  +\cos(  \pi \sigma  \rho) \right) + \frac{1}{2} &=0 \, , &&
\end{align}
where $\mathrm{sinc}(x) = \frac{\sin(x)}{x}$ is the standard \emph{sinus 
cardinal} function.  Let us introduce further the function
\begin{align}
\label{def:f}
f_{\sigma,T}(\rho)& =   \frac{\sigma^2}{2 T} \mathrm{\sinc} \left( \pi \sigma \rho \right)  +\cos \left(\pi \sigma  \rho \right) \, , &&
\end{align}
plotted in Fig. \ref{fig:f_and_rhomax} (Left). 

\begin{figure}
     \centering
         \includegraphics[width= 0.65\textwidth]{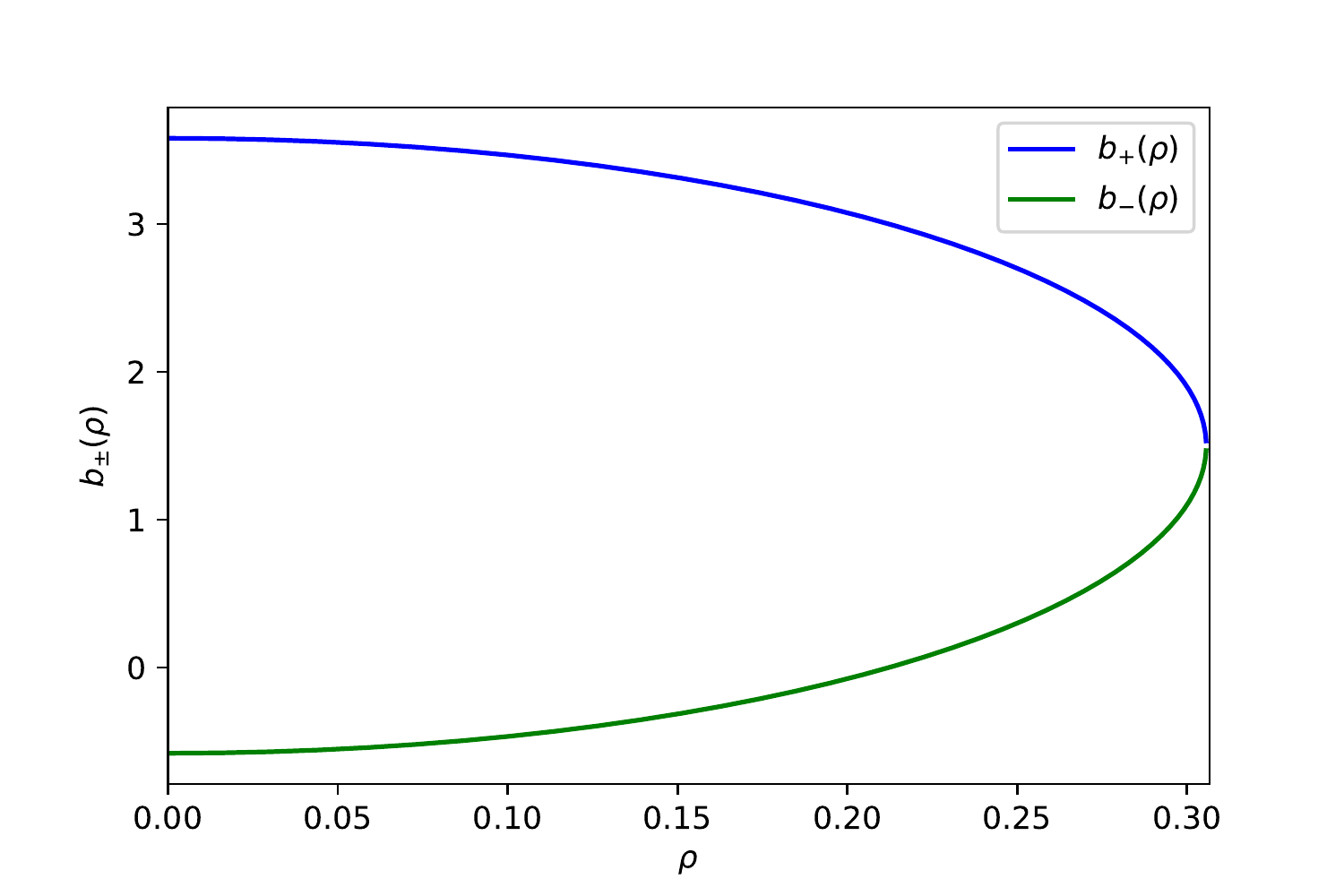}
         \label{fig:bofrho}
     \caption{Plot of the two inverse functions $b_{\pm}(\rho)$ given in Eq. \eqref{eq:paramsol.0}, for 
$\sigma = T =1$. Rotating the figure anticlockwise, followed by
a reflection around the vertical axis, gives $\rho_T(b)$ as
a function of $b$, as shown in Fig. \ref{fig:DBM_and_density} (Right).  }
     \label{fig:parametricsol}
\end{figure}

Let us now imagine that the value of $\rho$ is fixed. Then the two 
solutions $w_{\pm}(\rho)$ of the system 
(\ref{eq:SecondOrderw}) are given in terms of this 
function $f_{\sigma,T}(\rho)$ by,
\begin{align}
\label{prop:solution_w}
w_{\pm}(\rho) &= f_{\sigma,T}(\rho)  
\pm \sqrt{ f_{\sigma,T}( \rho)^2 -1 } \, , &&
\end{align}
and they satisfy the symmetry relation 
\begin{align}
\label{prop:symmetry_relation}
\frac{1}{w_{-}(\rho)} &= w_{+}(\rho) \, . &&
\end{align}
Injecting this into the top equation of (\ref{eq:RealandIm1}) 
we get two solutions $b_{\pm}(\rho)$: 
\begin{align}
b_{\pm}(\rho) &= 1 +  \frac{\sigma}{2} + \frac{T}{\sigma} 
\ln  w_{\pm}(\rho) + \frac{T}{2 \sigma \, 
\sinc ( \pi \sigma \rho)} \left( w_{\pm}- \frac{1}{w_{\pm}} 
\right) \, . &&
\label{eq:paramsol.0}
\end{align}
Using the symmetry relation \eqref{prop:symmetry_relation} and 
the expression \eqref{prop:solution_w} for $w_{\pm}(\rho)$, 
we get the following parametric solution for the density 
\begin{align}
\label{eq:parametricsolution}
b_{\pm}(\rho) &= 1+ \frac{\sigma}{2} \pm   
\frac{T}{\sigma} \left(   \ln \left( f_{\sigma,T}(\rho)+ \sqrt{f_{\sigma,T}(\rho)^2 -1} \right) + \frac{\sqrt{f_{\sigma,T}(\rho)^2 -1}}{\sinc \left( \pi \sigma \rho \right) } \right) \, .&&
\end{align}
In Fig. \ref{fig:parametricsol}, we plot the two branches $b_{\pm}(\rho)$
as a function of $\rho$ for fixed $T$. Indeed, if one rotates this
plot anticlockwise by $\pi/2$ and then reflects around the vertical axis,
one gets the desired density $\rho_T(b)$ as a function of $b$, as seen
in Fig. \ref{fig:DBM_and_density} (Right). Apart from being able
to plot the density, one can also extract a few additional
information from the explicit
expression in Eq. (\ref{eq:parametricsolution}), as discussed below.

\vskip 0.3cm

\noindent{\emph {Maximum value of the density:}} From Eq. (\ref{eq:parametricsolution}), 
we see that the maximum value of the density $\rho^{\star}$  
is attained at the point $b^{\star}$ for which $b_{+}(\rho) = b_{-}(\rho)$,
i.e., $b^{\star}=1 + \frac{\sigma}{2}$. 
The value of the maximum of the density $\rho^{\star} = 
\rho_T \left(b^{\star} \right)$, is therefore given as 
the \emph{first positive} solution of 
\begin{align}
f_{\sigma,T}(\rho^{\star}) &= 1 \, ,&&
\label{eq:rhomax.1}
\end{align}
which using Eq. \eqref{def:f} is equivalent to finding 
the first positive solution of
\begin{align}
\rho^{\star} \tan \left( \frac{\sigma \pi \rho^{\star} }{2} \right) 
&= \frac{\sigma }{2 T \pi} \, .&&
\label{eq:rhomax.2}
\end{align}
A plot of the maximum $\rho^{\star}$ as a function of $T$  
for $\sigma =1$ is given in Fig. \ref{fig:f_and_rhomax} (Right).

\vskip 0.3cm 

\noindent {\emph {Behavior near the edges:}} 
By Taylor expanding the function $b_{\pm}(\rho) $ near $\rho=0$, we have 
\begin{align}
\label{brho_small}
b_{\pm}(\rho) & \sim b_{\pm}(T) \mp 
\frac{T \pi^2 }{2} \sqrt{\sigma^2+4T}\, \rho^2 \quad \mbox{ as } \rho \to 0 \, ,&&
\end{align}
where the edges $b_{\pm}(T)$ are given by
\begin{align}
b_{\pm}(T) &= 1+\frac{\sigma}{2}\pm 
\left[\frac{\sigma}{2}\, \sqrt{1+\frac{4T}{\sigma^2}}+ \frac{T}{\sigma}\,
\ln \left(1+ \frac{\sigma^2}{2T}+ \frac{\sigma^2}{2T}\,
\sqrt{1+ \frac{4T}{\sigma^2}}\right)\right]\, .  &&
\label{b_edge.1}
\end{align}
It is easy to verify that the expression for $b_{-}(T)$
coincides with Eq. (\ref{bl_flat.1}).
Inverting the relation in Eq. (\ref{brho_small}), 
one finds that the density vanishes as a square root near the edge, 
with a prefactor that can be computed explicitly 
\begin{align}
\rho_T(b)  &\sim  
\frac{1}{\pi}  \sqrt{\frac{2}{T}} 
\frac{1}{(\sigma^2+4T)^{\frac{1}{4}}}  \sqrt{ (b_{\pm}(T) - b )_+} \quad \mbox{ as } b \to b_{\pm}(T) \, .&&
\label{eq:behavioredge}
\end{align}
where $(x)_+$ is equal to $x$ for $x>0$ and $0$ otherwise.

\section{The deformed GUE with flat initial condition and the left large deviation function of the dynamical system}
\label{sec:DeformedGUE_&_LDP}

So far, we have computed the average density of eigenvalues 
$\rho_T(b)$ in the large $N$
limit of the relaxation matrix, $\mat{B}=\mat{A}+\sqrt{T}\, 
\mat{J}$ for any $T$, where $J$ is an $N\times N$ GOE matrix and
$\mat{A}$ is diagonal with positive entries drawn from a flat
distribution over $[1,1+\sigma]$ with width $\sigma$. This
gives us the exact $T_c$ between the stable to unstable transition.
We expect that for finite but large $N$, the probability of stability
will have a qualitatively similar behavior as in the
homogeneous model in Eq. (\ref{summary_stable.1}), see also Fig. \ref{fig:PvsT}:

\begin{align}
\label{summary_stable.2}
     \mathcal{P}_{\mathrm{stable}}(T,N) &\approx \left\{
    \begin{array}{lll}
        \mathrm{exp} \left[ - \frac{N^2}{2} \Phi_{+}(\sigma,T) +o(N^2) \right]
& \mbox{for } T > T_c  \mbox{ and } |T-T_c| \sim O(1)  \, ,\\
\\
        \mathcal{F}^{(1)}\left( \gamma N^{2/3}\left(T^{-1/2}-T_c^{-1/2}\right) \right) &
\mbox{for } |T -T_c| \sim O(N^{-\frac{2}{3}}) \, , \\
\\
        1 -\mathrm{exp} \left[ - \frac{N}{2} \Phi_{-}(\sigma,T) +o(N)  \right] &
\mbox{for } T<T_c \mbox{ and } |T-T_c| \sim O(1) \, ,
    \end{array}
\right. &&
 \end{align}
where $\gamma$ is a constant of order one and  and the large deviation functions $\Phi_{\pm}(\sigma,T)$ on
either sides of $T_c$ would be different. Note that for $\sigma=0$, the large deviation functions $\Phi_{\pm}(T) = \Phi_{\pm}(\sigma=0,T)$ could be related to the cumulative distribution of the top eigenvalue $\lambda_N$ of the matrix $\mathbf{J}$, see Eq. \eqref{eq:PhiPsi}. However for $\sigma>0$, there is no such relation and one needs to compute $\Phi_{\pm}(\sigma,T)$ from first principles.  It turns out (see later) that to compute
the large deviation functions $\Phi_{\pm}(\sigma,T)$,
we need the information
on the full joint distribution of eigenvalues, and not just
the one point function, i.e., the average density. 

Hence, our next natural step was to see if we could compute the joint 
distribution of the eigenvalues of $\mat{B}$, where $\mat{J}$
is a GOE matrix. Unfortunately, we did not
succeed yet in that. However, it turns out that one can compute
the joint distribution of eigenvalues in a Hermitian counterpart
of the relaxation matrix, 
$\tilde{\mat{B}} = \mat{A} + \sqrt{T} \tilde{\mat{J}}$,
where $\mat{A}$ is still diagonal with a flat distribution, but
now $\mat{J}$ is Hermitian, i.e., a GUE matrix. 

In this section, for this deformed GUE model, 
we derive an explicit formula for the 
joint law of eigenvalues for the flat initial condition, thanks to the 
Itzykson-Zuber determinantal formula.  We will see that this
leads to a new Coulomb gas, where the eigenvalues can be interpreted as the positions of a gas of particles confined in a harmonic potential and repelling pairwise as in the standard GUE, but with an additional twist that the pairwise interaction here is a \emph{linear combination} of a logarithmic (as in standard GUE) and a log-sinh type interaction. Finally, using this Hermitian modification, we will show how to compute at least the large deviation function 
$\Phi_{-}(\sigma,T)$ appearing in Eq. \eqref{summary_stable.2}, in  the `weakly stable' phase ($T <T_c$) 
in the original deformed GOE model. However, computing the large deviation function
$\Phi_{+}(\sigma,T)$ on the `strongly unstable' phase ($T>T_c$) 
still remains out of reach.

\subsection{The deformed GUE with flat initial condition and its joint law for the eigenvalues}
\label{sec:deformedGUE}

The deformed GUE model   
\cite{o2001brownian,baryshnikov2001gues,noack2020concentration,krajenbrink2021tilted} 
is the Hermitian counterpart of the deformed GOE model 
\begin{align}
\label{def:deformedGUE}
\tilde{\mat{B}} &= \mat{A} + \sqrt{T} \tilde{\mat{J}} \, , &&
\end{align}
with the matrix $\mathbf{A}= \Diag(a_1, \dots, a_N)$ with positive entries
as before. The matrix $\tilde{\mat{J}}$ is a GUE matrix 
whose law is given by: 
\begin{align}
\label{def:GUE}
\mathcal{P}_N (\tilde{\mat{J}}) \mathrm{d}\tilde{\mat{J}} &= \mathrm{exp} \left[ - N \Tr \frac{\tilde{\mat{J}}^2}{2} \right] \mathrm{d}\tilde{\mat{J}} \, , &&
\end{align}
with now $\mathrm{d}\tilde{\mat{J}} = \prod_{i=1}^N 
\mathrm{d}\tilde{J}_{ii} \prod_{j>i} \mathrm{d} \mathfrak{Re} 
\tilde{J}_{ij} \, \mathrm{d} \mathfrak{Im} \tilde{J}_{ij}$, the Lebesgue 
measure on the space of Hermitian matrices. One can repeat the second 
perturbation theory analysis for the eigenvalues and shows that they 
follow the $\beta =2$ DBM \eqref{DBM}, with diffusion
constant $D=1/(2N)$. In the large $N$ limit, we get 
the same Burgers' equation \eqref{def:BurgersEq} for the resolvent  and 
so one gets the same limiting density\footnote{note that the exponent in 
the exponential function in Eq. \eqref{def:GUE} differed by a factor 
two from Eq. \eqref{JL_GOE} to have exactly the same limiting spectral 
distribution.} $\rho_T(b)$.
Hence, both the deformed GOE and
the deformed GUE share the same limiting density $\rho_T(b)$
for arbitrary initial density $\mu(a)$ of the $a_i$'s.  
In particular, for
the flat initial condition, this common density is 
given by the parametric solution of Eq.
\eqref{eq:parametricsolution}.

From the probability density in Eq. \eqref{def:GUE} of the matrix 
$\tilde{\mat{J}}$, as it was done in Eq. (\ref{eq:prob_of_B}),
one can easily obtain the distribution 
of the matrix $\tilde{\mat{B}}$, using the relation
 \eqref{def:deformedGUE}. One gets 
\begin{align}
\label{prob_B.1}
\mathcal{P}_N\left(\tilde{\mat{B}} \right)\mathrm{d} 
\tilde{\mat{B}} &\propto \mathrm{exp} \left[- \frac{N}{T} \Tr   
\frac{\left(  \mat{A} - \tilde{\mat{B}} \right)^2 }{2} 
\right] \mathrm{d}\tilde{\mat{B}} \, , &&
\end{align}
with $\mathrm{d}\tilde{\mat{B}} = \prod_{i=1}^N \mathrm{d}\tilde{B}_{ii} 
\prod_{1 \leq i \leq j \leq N} \mathrm{d} \mathfrak{Re} \tilde{B}_{ij} 
\mathrm{d} \mathfrak{Im} \tilde{B}_{ij}$. Since $\tilde{\mat B}$ is 
Hermitian, it admits a spectral decomposition $\tilde{\mat{B}}=\mat{U} 
\Diag(\tilde{b}_1,\dots, \tilde{b}_N) \mat{U}^*$, where $\mat{U}$ is a 
random unitary matrix. By expanding the square and integrating over the 
group $\mathsf{U}(N)$ of unitary matrix, we have for the joint law of 
the eigenvalues:
\begin{align}
\label{JL_Hermitian}
\mathcal{P}_N( \tilde{b}_1, \dots,  \tilde{b}_N)  &\propto e^{- \frac{N}{2 T} \sum_{i=1}^N \tilde{b}_i^2} \Delta( \tilde{b}_1, \dots  \tilde{b}_N)^2 \left( \int_{ \mathsf{U}\left(N\right)}  \,   e^{  \frac{N}{ T} \Tr \left( \mat{A} \, \mat U \,  \Diag( \tilde{b}_1, \dots, \tilde{b}_N) \, \mat U^* \right)  }  \mathrm{d} \mat U \right)  \, , &&
\end{align}
where $\Delta(\tilde{b}_1, \dots \tilde{b}_N) = \prod_{i <j} (\tilde{b}_j 
- \tilde{b}_i)= \det \left[ \tilde{b}_j^{i-1}\right]_{1 \leq i,j\leq N } $ 
is the \emph{Vandermonde product} that appears in the Jacobian 
of the change of variables 
$\tilde{\mat{B}} \to \left( (\tilde{b}_1, \dots, \tilde{b}_N) , 
\mat{U} \right)$. For convenience, let us first order the
eigenvalues such that $\tilde{b}_1< \tilde{b}_2<\ldots <\tilde{b}_N$
and similarly $a_1<a_2<\ldots< a_N$. We can unorder them later. 
The integral over the Haar group in
Eq. (\ref{JL_Hermitian}) can be identified,
up to a trivial factor $\frac{1}{T}$ inside the exponential, 
to the well known $\beta =2$ \emph{Harish-Chandra-Itzykson-Zuber} (HCIZ) 
integral \cite{harish1956invariant,harish1957differential,itzykson1980planar},  
\begin{align}
\label{def:HCIZ_b2}
 \mathcal{I}^{(2)}_N\left( \mat{A},  \tilde{\mat{B}}\right) &= \int_{ \mathsf{U}\left(N\right)}   \,   e^{  N \Tr \left( \mat{A} \, \mat U \,  \tilde{\mat{B}} \, \mat U^* \right)  } \mathrm{d} \mat U \,.&&
\end{align}
This integral can be explicitly carried out giving the
beautiful Itzykson-Zuber determinantal formula \cite{itzykson1980planar}  
\begin{align}
\label{eq:IZformula}
\mathcal{I}^{(2)}_N\left( \mat{A},  \tilde{\mat{B}}\right) &= 
\frac{\prod_{l=1}^{N-1} l!}{N^{\frac{N^2 -N}{2}}} 
\frac{\det \left[ \mathrm{e}^{N a_i \tilde{b}_j} 
\right]_{1 \leq i,j\leq N } }{\Delta(a_1, \dots a_N)  
\Delta(\tilde{b}_1, \dots, \tilde{b}_N) } \, . &&
\end{align}
By injecting this expression into Eq. \eqref{JL_Hermitian} 
and only keeping the terms depending on the  $\tilde{b}_i$, 
we have for the joint law of the eigenvalues, 
\begin{align}
\label{eq:JL_dGUE_genericA}
\mathcal{P}_N( \tilde{b}_1, \dots, \tilde{b}_N )  &\propto  e^{- \frac{N}{2T} \sum_{i=1}^N \tilde{b}_i^2} \Delta(\tilde{b}_1, \dots, \tilde{b}_N) \det \left[ e^{\frac{N}{T} a_i \tilde{b}_j} \right]_{1 \leq i,j\leq N } \, . &&
\end{align}
At this stage, the equation \eqref{eq:JL_dGUE_genericA} 
holds for an arbitrary diagonal matrix 
$\mathbf{A}=\Diag(a_1, \dots, a_N)$ with $a_i$'s ordered. 
Let us now take the $a_i$'s 
to be given by the flat initial 
condition (\ref{flatIC}). In this case, the determinant appearing 
in Eq. \eqref{eq:JL_dGUE_genericA} considerably simplifies since  
\begin{align}
\det \left[ e^{ \left( \frac{N}{T} + \frac{\sigma (i-1)}{T} \right) 
\tilde{b}_j} \right] &= \mathrm{e}^{\frac{N}{T} \sum_{i=1}^N\tilde{b}_i} 
\Delta \left( e^{ \frac{\sigma}{T}\tilde{b}_1}, \dots, 
e^{ \frac{\sigma}{T}\tilde{b}_N}  \right) = 
\mathrm{e}^{\frac{N}{T} \sum_{i=1}^N \tilde{b}_i} 
\prod_{ i < j} \left(e^{\frac{\sigma}{T} \tilde{b}_j } - 
e^{\frac{\sigma}{T}\tilde{b}_i}\right) \, . &&
\end{align}
Hence, the joint law for the ordered eigenvalues (\ref{eq:JL_dGUE_genericA})
simplifies to 
\begin{align}
\label{eq:JointDistributionFlat}
\mathcal{P}_N( \tilde{b}_1, \dots, \tilde{b}_N )  
&\propto \mathrm{exp} \left[ \sum_{i=1}^N \frac{N}{T} 
\left(-\frac{ \tilde{b}_i^2}{2} + \tilde{b}_i \right) 
\right] \Delta(\tilde{b}_1, \dots, \tilde{b}_N) \, 
\Delta \left( e^{ \frac{\sigma}{T}\tilde{b}_1}, \dots, 
e^{ \frac{\sigma}{T}\tilde{b}_N}  \right) \,.&&
\end{align}
Note that if the eigenvalues $\tilde{b}_i$'s are now
unordered, their joint distribution just reads
\begin{align}
\label{eq:JointDistributionFlat_unordered}
\mathcal{P}_N( \tilde{b}_1, \dots, \tilde{b}_N )
&\propto \mathrm{exp} \left[ \sum_{i=1}^N \frac{N}{T}
\left(-\frac{ \tilde{b}_i^2}{2} + \tilde{b}_i \right)
\right] \left|\Delta(\tilde{b}_1, \dots, \tilde{b}_N)\right| \,
\left|\Delta \left( e^{ \frac{\sigma}{T}\tilde{b}_1}, \dots,
e^{ \frac{\sigma}{T}\tilde{b}_N}  \right)\right| \,.&&
\end{align}
Using the identity 
\begin{align}
\left( \mathrm{e}^{x} - \mathrm{e}^{y} \right) 
\mathrm{e}^{ -\frac{(x+y)}{2}}&=2\,\sinh \left( \frac{x-y}{2} \right)\, , &&
\label{iden_sinh}
\end{align}
we can write the second Vandermonde in 
Eq. (\ref{eq:JointDistributionFlat_unordered}) as 
\begin{align}
\label{vdm.1}
\left|\Delta \left( e^{ \frac{\sigma}{T}\tilde{b}_1}, \dots, 
e^{ \frac{\sigma}{T}\tilde{b}_N}  \right)\right| & \propto \mathrm{exp} \left[  
\left( \frac{\sigma}{2T}  \sum_{i\neq j} (b_i+b_j) \right) + 
\frac{1}{2} \sum_{i\neq j} \ln \sinh \left( \frac{\sigma}{2T} | b_i - b_j | 
\right) \right] \, , && \nonumber \\
&= \mathrm{exp} \left[ \frac{\sigma (N-1)}{2T}  
\sum_{i} b_i + \frac{1}{2} \sum_{i\neq j} 
\ln \sinh \left( \frac{\sigma}{2T} | b_i - b_j | \right)  \right] \, .&&
\end{align}
Using Eq. \eqref{eq:JointDistributionFlat_unordered} and completing the square, 
this can be finally written as
\begin{align}
\label{prop:JL_logsinh}
\mathcal{P}_N (\tilde{b}_1, \dots, \tilde{b}_N )  
&\propto\mathrm{exp} 
\left[ - N \sum_{i=1}^N \frac{\left( \tilde{b}_i 
-\left( 1+ \frac{\sigma}{2}  
\frac{N -1}{N} \right) \right)^2}{2 T}  +  
\frac{1}{2} \sum_{i \neq j} 
\ln | \tilde{b}_i - \tilde{b}_j | +  
\frac{1}{2}  \sum_{i \neq j} \ln  \sinh 
\left( \frac{\sigma}{2 T} | \tilde{b}_i - \tilde{b}_j | \right) 
\right] \, .&&
\end{align}
Eq. (\ref{prop:JL_logsinh}) provides a nice Coulomb gas interpretation
of the joint law of eigenvalues.  
The joint distribution in Eq. (\ref{prop:JL_logsinh}) can be
written as a Boltzmann distribution $\sim e^{-E(\{\tilde{b}_i\})}$,
where the energy function can be read off the argument of
the exponential in Eq. (\ref{prop:JL_logsinh}).
The eigenvalues $\{\tilde{b}_i\}$'s can be interpreted as 
the positions of $N$ charges on a line. These charges
are subjected to an external harmonic potential centered at
$\tilde{b}= 1+ (\sigma/2)(N-1)/N$. In addition, they
repel each other pairwise. The pairwise interaction
is a linear combination of the logarithmic repulsion
(represented by the second term inside the exponential
in Eq. (\ref{prop:JL_logsinh})) and a log-sinh
interaction (the third term in Eq. (\ref{prop:JL_logsinh})).
In the limit $\sigma\to 0$ (upon absorbing an overall constant
in the normalization), the third term also becomes logarithmic,
and hence the system reduces to the standard log-gas of Gaussian
random matrices \cite{forrester2010log}. 
But for a nonzero $\sigma>0$, we have a new
variety of Coulomb gas with both log and log-sinh interactions
that is usually not encountered in RMT models.

Given the joint density of the eigenvalues in the Coulomb gas
representation in Eq. (\ref{prop:JL_logsinh}), one can, in principle,
obtain the average density in the large $N$ limit by a variational principle,
i.e., by employing a saddle point method for large $N$ to evaluate
the partition function of the Coulomb gas. This amounts to
minimizing the energy function $E(\{\tilde{b}_i\})$. Minimizing
this energy in Eq. (\ref{prop:JL_logsinh}) gives the saddle
point equation
\begin{align}
\label{eq:SaddlePoint}
\frac{1}{T} \left( 1+ \frac{\sigma}{2} \frac{N-1}{N} - \tilde{b}\right) +  \frac{1}{N}\sum_{j:j\neq i} 
\frac{1}{\tilde{b}_i -\tilde{b}_j} + 
\frac{\sigma}{T} \frac{1}{2N}\, 
\sum_{j:j\neq i} \coth\left(\frac{\sigma}{2T}(\tilde{b}_i-
\tilde{b}_j)\right) &=0 \, .&&
\end{align}
For large $N$, the sums can be replaced by integrals, and one obtains
an integral equation satisfied by the density $\rho_T(\tilde{b})$
\begin{align}
\label{eq:SP_density}
 \frac{1}{T}( b^{\star} - \tilde{b}) + \, {\rm Pr} \int
\frac{\rho_T(\tilde{y})\, \mathrm{d}\tilde{y}}{\tilde{b}^*-\tilde{y}}
+ \frac{\sigma}{2T} {\rm Pr} \int \rho_T(\tilde{y})\, 
\coth\left(\frac{\sigma}{2T}(\tilde{b}-\tilde{y})\right)\, \mathrm{d}\tilde{y}&=0\, , &&
\end{align}
where we recall $b^{\star} = 1 + \sigma/2$, ${\rm Pr}$ denotes the principal value and the integral
equation holds for all $\tilde{b}\in [\tilde{b}_{-}(T), \tilde{b}_{+}(T)]$
where $\tilde{b}_{\pm}(T)$ denotes the support edges. On physical grounds,
we expect the
density to have only a single support on the real line.  

In the limit $\sigma\to 0$, the third term coincides with the second term
in Eq. (\ref{eq:SP_density}), and one recovers the standard saddle
point density of the log-gas \cite{forrester2010log,majumdar2014top}, 
\begin{align}
\label{SP_loggas}
\frac{1}{2T}( b^{\star}- \tilde{b}) + \, {\rm Pr} \int
\frac{\rho_T(\tilde{y})\, \mathrm{d}\tilde{y}}{\tilde{b}-\tilde{y}}&=0\, . &&
\end{align}
This singular value integral equation can be inverted using
Tricomi's formula (see Ref. \cite{majumdar2014top} for details)
and one recovers the shifted semi-circular law in 
Eq. (\ref{shifted_Wigner.1}). For a nonzero $\sigma$,
we were not able to solve the singular integral equation 
(\ref{eq:SP_density}). However, remarkably, we actually know
the solution $\rho_T(\tilde{b})$, albeit in a parametric form,
in Eq. (\ref{eq:parametricsolution}) via the resolvent method.
Note that the parametric solution in Eq. (\ref{eq:parametricsolution})
also holds for deformed GUE $\rho_T(\tilde{b})$ which is
identical to that of deformed GOE, as shown earlier.
It then remains a mathematical challenge to derive this
parametric solution (\ref{eq:parametricsolution}) directly
from the singular value integral equation (\ref{eq:SP_density}).

\subsection{Relations to other models}
\label{sec:othermodels}

The matrix $\tilde{\mat{B}}$ (and the matrix $\mat{B}$ of the original 
model) as described in the previous section is related to several models
of RMT that have appeared before in the literature. The joint density for the 
matrix $\tilde{\mat{B}}$ in Eq. (\ref{prob_B.1})
can be written, upon absorbing the $\Tr (\mat{A}^2)$ in the normalization constant, as
\begin{align}
\label{def:ExternalSource}
\mathcal{P}_N (\tilde{\mat{B}})\mathrm{d}\tilde{\mat{B}} &\propto e^{- N \Tr \left[ V(\tilde{\mat B}) - \tilde{\mat{A}} \tilde{\mat{B}}  \right] } \mathrm{d}\tilde{\mat{B}} \, , &&
\end{align}
with $V(x)= \frac{x^2}{2}$ and  $\tilde{\mat{A}} = \frac{\mat{A}}{T}$. The matrix $\tilde{\mat{A}}$ in Eq. \eqref{def:ExternalSource} plays the role of an external field, and hence models of the type \eqref{def:ExternalSource} are known as  random matrices with an external source \cite{brezin2016random}. A particular interest has been devoted to the case where one half of the eigenvalues of the matrix  $\tilde{\mat{A}}$ takes the value $a$ and the other half takes the value $-a$, see \cite{brezin1998level,bleher2004large,aptekarev2005large,bleher2007large}. The local properties for the  case of flat initial condition \eqref{flatIC} has also been studied in \cite{claeys2014random} using Riemann-Hilbert techniques. 
 
From Eq. \eqref{eq:JointDistributionFlat}, one can see that the 
joint law of eigenvalues exhibits a bi-orthogonal structure of a 
determinantal point process which resembles somewhat the 
Muttalib-Borodin ensemble with parameter $\theta>0$ 
\cite{muttalib1995random,borodin1998biorthogonal} 
\begin{align}
\mathcal{P}_N( \lambda_1, \dots, \lambda_N )  &\propto \mathrm{exp} \left[ -N  \sum_{i=1}^N V(\lambda_i) \right] \Delta(\lambda_1, \dots, \lambda_N) \, \Delta \left( \lambda_1^{\theta}, \dots,  \lambda_N^{\theta}  \right) \, .&&
\label{def:MuttalibBorodin}
\end{align}
with the difference that in the second Vandermonde, 
the arguments are exponential in Eq. \eqref{eq:JointDistributionFlat}, 
while they have a power-law form in Eq. \eqref{def:MuttalibBorodin}. However, the case with the exponential function in the second Vandermonde, appeared in the randomized multiplicative Horn problem \cite{zhang2019harmonic}, in the DPMK equation for transport in semiconductors \cite{beenakker1997random}
and in the multiplicative analogue of Dyson Brownian Motion \cite{ipsen2016isotropic}. \\

If one makes the change of variable 
$ x_i = \frac{1}{\sqrt{T}}  
\left(\tilde{b}_i - (1+\frac{\sigma}{2}) \frac{N-1}{N} \right)$ in
Eq. (\ref{prop:JL_logsinh}) and writes 
$ r= \frac{\sigma}{2 \sqrt{T}}$, the joint distribution of the 
$x_i$'s is given by: 
\begin{align}
\label{def:rescaledjointlaw}
\mathcal{P}_N( x_1, \dots, x_N )  &\propto 
\mathrm{exp}  \left[- \frac{N}{2 } \sum_{i=1}^N x_i^2 + 
\frac{1}{2} \sum_{i \neq j} \ln | x_j - x_i | +
\frac{1}{2} \sum_{i \neq j} \ln  \mathrm{sinh}  
\left( r | x_j - x_i | \right) \right] \, .  &&
\end{align}
Thus, we have a Coulomb gas where the pairwise interaction is 
a linear combination of logarithmic and log-sinh. The case with
only log-repulsion (without the log-sinh) corresponds to the standard Gaussian matrices.
The case with only log-sinh repulsion (without the log term) appears
in the partition function of the 
Chern-Simons model on $S^3$ \cite{marino2005chern,marino2006matrix}, in the theory of Stietljes-Wigert polynomials 
\cite{dolivet2007chern,tierz2010schur,szabo2010chern,takahashi2012noncolliding} and 
in the recent study of vicious walkers constrained at both ends by a 
flat initial conditions \cite{grela2021non}. The parameter 
$r=\sigma/\sqrt{4T}$ in Eq. 
\eqref{def:rescaledjointlaw} controls the strength of the second 
interaction, since for $r$ positive, the function $\ln \sinh(r)$ is 
increasing from $0$ to $\infty$. In the limit
$r\to 0$, it reduces to the log-gas as shown before.
In the opposite limit $r\to \infty$, Eq. (\ref{def:rescaledjointlaw})
to leading order in $r$ reduces to a $1$D-one component plasma (OCP)
model \cite{lenard1961exact,prager1962one,baxter_1963} 
\begin{align}
    \mathcal{P}_N( x_1, \dots, x_N )  &\propto \mathrm{exp}  
\left[- \frac{N}{2 } \sum_{i=1}^N x_i^2 + 
\frac{1}{2} \sum_{i \neq j}| x_j - x_i |\right] \, , &&
\label{eq:1d2dmix}
\end{align}
for which the equilibrium measure is the flat distribution and 
the distribution of its largest (lowest) eigenvalue
have recently been computed exactly, both
for typical fluctuations and also for large deviations
\cite{dhar2017exact,dhar2018extreme,flack2021truncated}.

\subsection{Large deviation below the critical strength  \texorpdfstring{$T_c$}{Tc}  for the flat initial condition}
\label{sec:LDP}

We now go back to the original deformed GOE model with flat initial 
condition \eqref{flatIC}. In the strict $N\to \infty$ limit,
the probability of stability $\mathcal{P}_{\mathrm{stable}}(N\to \infty,T)$
follows the step function behavior as in Eq. \eqref{prop:ProbaVST}.
We have computed the exact $T_c$ and also
the average density of particles in 
a parametric form \eqref{eq:parametricsolution} for 
the flat initial condition (\ref{flatIC}). As we have discussed in the 
introduction, the next step it to derive the behavior of the 
probability $\mathcal{P}_{\mathrm{stable}}(N,T)$ for large but finite 
$N$, close to the critical point $T=T_c$. 
Similar to May's original homogeneous model in Eq. (\ref{summary_stable.1}), 
one can show 
\cite{lee2015edge} that the typical `small' fluctuations of $O(N^{-2/3})$
around $T=T_c$, are again described by the 
Tracy-Widom distribution. This is the middle equation of Eq. \eqref{summary_stable.2} where the constant $\gamma$  in Eq. \eqref{summary_stable.2} is given in \cite{lee2015edge}.  For $\sigma >0$, The large deviation
functions $\Phi_{\pm}(\sigma,T)$ are expected
to be different from the homogeneous model $ \Phi_{\pm}(\sigma =0,T)= \Phi_{\pm}(T) = \Psi_{\mp} \left( \frac{1}{\sqrt{T}} \right)$, with $ \Psi_{\mp}$  given in Eqs \eqref{left.1} and \eqref{right.1}.
For values of $T>T_c$ (see Fig. 
\ref{fig:DBM_and_density} (Left)) a finite fraction of the 
eigenvalues is negative and as explained in the introduction, to access 
the large deviation regime one needs to push all those eigenvalues leading
to a modification of the equilibrium density in the bulk. For the matrix 
$\mathbf{B}$, the eigenvalues do not behave as a simple $2$D Coulomb-gas 
particles confined on the real line and therefore this equilibrium 
density in the presence of a pushing wall, 
needed for the computation of the large deviation 
function $\Phi_{\pm}(\sigma,T)$ in this regime, is hard to obtain. 
For this reason, we restrict 
the discussion only to the weakly stable phase, corresponding to $T<T_c$, where 
the bulk density remains unchanged when one pulls
a single charge out of the bulk
and is still given by $\rho_T$. To access the 
large deviation function $\Phi_{-}(\sigma,T)$ in this regime, 
we recall using Eq. 
\eqref{crit_stability} that one has to compute

\begin{align}
    \mathcal{P}_{\mathrm{stable}}(T,N) &= 1- \mathrm{Prob}\left[ b_1 <0 \right] \,, &&
    \label{eq:probstab.finalsec}
\end{align}
where to ease notation, we simply write $b_i \equiv b_i(T)$ in the rest of this section and the eigenvalues $\{ b_i\}$ are in increasing order. To evaluate this probability, we will redo a similar computation as the one in Sec. \ref{sec:deformedGUE} to obtain the joint law for the eigenvalues $(b_1, \dots, b_N)$. The main difference with the Hermitian case is that the joint law will involve the $\beta=1$ HCIZ integral. Instead of the $\beta=2$ HCIZ integral, there is no simple determinantal formula for the $\beta=1$ case. It will be possible to overcome this difficulty thanks to the known asymptotic of the HCIZ integral, and we can then compute the probability by integrating the joint law over all eigenvalues and the use of a standard saddle-point approximation.  
\vskip 0.3 cm

\noindent From the law of the matrix elements \eqref{eq:prob_of_B}, one 
obtains the joint law of the eigenvalues by the change of variable $ \left( \mat{B} \to (b_1, \dots, b_N), \mat{O} \right)$, where $\mat{O}$ is the orthogonal matrix of eigenvectors. This change of variable introduces a Vandermonde (but without the square), such that the joint law of eigenvalues can be written as:
\begin{align}
\label{eq:JL_initial}
\mathcal{P}_N( b_1, \dots, b_N)  &\propto e^{- \frac{N}{2 T} \sum_{i=1}^N \frac{b_i^2}{2}} \Delta(b_1, \dots b_N) \left( \int_{ \mathsf{O}\left(N\right)} \,   e^{  \frac{N}{ 2T} \Tr \left( \mat{A} \, \mat O \,  \Diag(b_1, \dots, b_N) \, \mat O^* \right)  }  \mathrm{d} \mat O   \right) \, . &&
\end{align}
This expression involves an integral of the form:
\begin{align}
 \mathcal{I}^{(1)}_N\left( \mat{A},  \mat{B}\right) &= \int_{ \mathsf{O}\left(N\right)}   \,   e^{  \frac{N}{2} \Tr \left( \mat{A} \, \mat O \,  \mat{B} \, \mat O^* \right)  } \mathrm{d} \mat O \, ,&&
 \label{eq:HCIZbeta1}
\end{align}
which is called the $\beta=1$ HCIZ integral. There is no simple Itzykson-Zuber formula \eqref{eq:IZformula} in this case, but since we are interested in the large $N$ limit, what one only needs is the asymptotic behavior of this integral.  For large $N$ and $\beta=1,2$, this integral is known to behave as: 
\begin{align}
\label{prop:Matystin}
 \mathcal{I}^{(\beta)}_N\left( \mat{A},  \mat{B}\right) &\approx  \mathrm{exp} \left[ \frac{N^2 \beta }{2} \mathcal{F}(a_1, \dots, a_N ; b_1, \dots, b_N) +o(N^2) \right] \, , &&
\end{align}
where the function $\mathcal{F}(.)$ satisfies a complex variational 
principle that was first derived by Matytsin \cite{matytsin1994large} in 
the Hermitian case and extended to the symmetric case in 
\cite{bun2014instanton,guionnet2002large}. Note that there is an 
additional one-half factor in the symmetric case ($\beta=1$) compared to 
the Hermitian case ($\beta=2$), a result sometimes referred to as 
“Zuber-$\frac{1}{2}$” law, see \cite{zuber2008large}. 
The important point is that the function $\mathcal{F}(.)$ does
not depend on $\beta$ and the $\beta$ dependence appears just
as a prefactor of $\mathcal{F}(.)$ in Eq. (\ref{prop:Matystin}).
Thus, for $N$ 
large, using Eq. \eqref{eq:JL_initial} and Eq. \eqref{prop:Matystin} the 
joint law for the Deformed GOE is asymptotically given by:

\begin{align}
\label{eq:JoinLaw_HCIZb1}
    \mathcal{P}_N( b_1, \dots, b_N)  &\propto \mathrm{exp} \left[ - \frac{N^2}{2} \left( \frac{1}{N} \sum_{i=1}^N \frac{b_i^2}{2T} - \frac{1}{N^2} \sum_{j \neq i}^N \ln | b_i - b_j | - \mathcal{F} \left(\frac{a_1}{T}, \dots, \frac{a_N}{T}; b_1, \dots,b_N \right)  +o(1) \right) \right] \, . &&
\end{align}
Similarly for the deformed GUE, the joint law can be written as: 
\begin{align}
\label{eq:JointLaw_HCIZb2}
    \mathcal{P}_N( \tilde{b}_1, \dots, \tilde{b}_N)  &\propto \mathrm{exp} \left[ - N^2 \left( \frac{1}{N} \sum_{i=1}^N \frac{\tilde{b}_i^2}{2T} - \frac{1}{N^2} \sum_{j \neq i}^N \ln | \tilde{b}_i - \tilde{b}_j | - \mathcal{F} \left(\frac{a_1}{T}, \dots, \frac{a_N}{T}; \tilde{b}_1, \dots,\tilde{b}_N \right)  +o(1) \right) \right] \, . &&
\end{align}
Comparing  Eq. \eqref{prop:JL_logsinh} and Eq. \eqref{eq:JointLaw_HCIZb2}, one gets for the flat initial case \eqref{flatIC} that the function $\mathcal{F}$ is asymptotically given by:
\begin{align}
\label{eq:Matystinfunction_FIC}
    \mathcal{F} \left(\frac{a_1}{T}, \dots, \frac{a_N}{T};b_1, \dots,b_N \right) &\approx \sum_{i=1}^N \frac{b_i b^{\star}}{2T} + \frac{1}{2N^2} \sum_{j \neq i} \ln \sinh \left( \frac{\sigma}{2T} |b_i - b_j| \right) - \frac{1}{2N^2} \sum_{j \neq i}^N \ln  |b_i - b_j| +C+o(1) \, , &&
\end{align}
where $C$ is a constant independent of the $\{b_i \}$ and $b^{\star} = 1+  \frac{\sigma}{2}$. Discarding sub-leading term in $N$ in Eq. \eqref{eq:JoinLaw_HCIZb1} with $\mathcal{F}$ given in Eq. \eqref{eq:Matystinfunction_FIC}, one has \emph{asymptotically}:
\begin{align}
\label{prop:JL_sym_logsinh}
\mathcal{P}_N( b_1, \dots, b_N)  &\propto \mathrm{exp} \left[- \frac{N^2}{2} \left(\frac{1}{2NT} \sum_{i=1}^N \left(b_i  -b^{\star} \right)^2 - \frac{1}{2N^2} \sum_{j\neq i} \ln | b_i - b_j |- \frac{1}{2N^2} \sum_{j\neq i} \ln \sinh \frac{\sigma}{2T}| b_i - b_j | +o(1)  \right)   \right] \, .
\end{align}

\begin{figure}
     \centering
         \includegraphics[width= 0.65\textwidth]{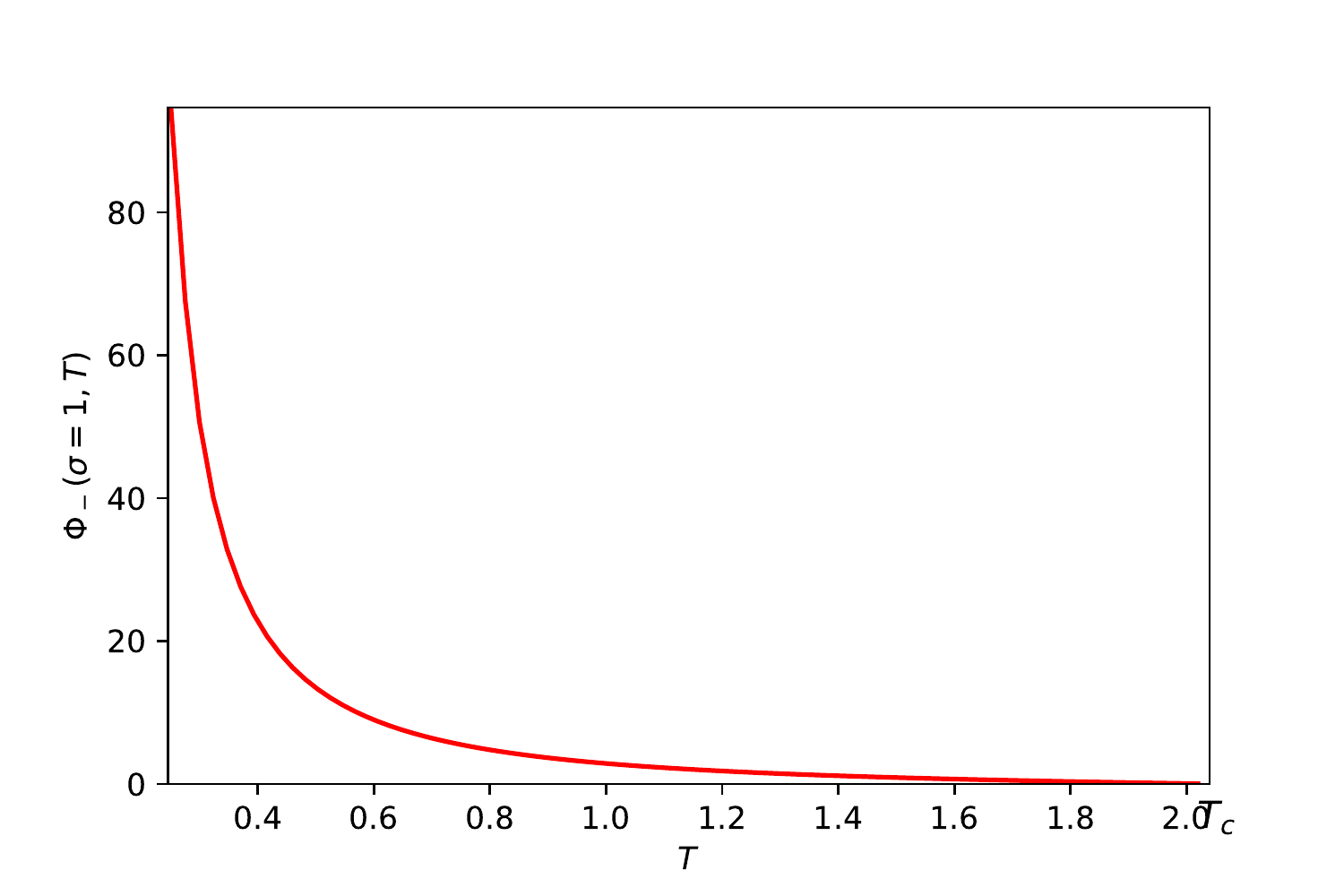}
         \label{fig:phiT}
     \caption{Plot of the rate function $\Phi_-(\sigma,T)$ defined by Eq. \eqref{eq:RateFunction} as a function of $T$, for $\sigma=1$.}
     \label{fig:LDplot}
\end{figure}

\vskip 0.3 cm
\noindent Now that we have the (asymptotic) behavior of the joint law pf the deformed GOE with flat initial condition, it is possible to compute the probability of stability of Eq. \eqref{eq:probstab.finalsec}. For the ordered eigenvalues, we can always express the cumulative probability as:
\begin{align}
    \mathrm{Prob} \left[ b_1 <0 \right] &= \int_{- \infty}^0 \left( \int_{- \infty}^{0} \dots \int_{- \infty}^{0}  \mathcal{P}_N(b_1, \dots, b_N) \mathrm{d}b_2 \dots \mathrm{d}b_N \right) \mathrm{d}b_1 \, . &&
    \label{eq:IntegralJointLaw}
\end{align}
Let us now separate the contribution given by respectively $b_1$ and by the other $N-1$ eigenvalues:
\begin{align}
\mathrm{Prob} \left[ b_1 <0 \right] &\approx \int_{- \infty}^0 \mathrm{e}^{-\frac{N}{4T} \left( b_1 - b^{\star} \right)^2} \, \left( \int_{- \infty}^{0} \dots \int_{- \infty}^{0}  \mathcal{P}_{N-1}(b_2, \dots, b_N) \mathrm{e}^{ \frac{1}{2}  \sum_{j=2}^{N} \ln (  b_j - b_1 ) + \frac{1}{2}  \ln \sinh  \frac{\sigma}{2T}(b_j - b_1)} \mathrm{d}b_2 \dots \mathrm{d}b_N \right) \mathrm{d}b_1 \, . &&
\label{eq:probab1.1}
\end{align}
In the weakly stable phase, the probability of having an unstable system corresponds to observe a rare configuration where the bottom eigenvalue $b_1$ is at position  far below its typical value $b_{-}(T)$ (which is above $0$ in the case $T <T_c$, see Fig. \ref{fig:DBM_and_density} (Left)). Since  $N$ is large, one may again expect that just moving one (the bottom) eigenvalue does not change the bulk density.  The empirical distribution of the other  $N-1$ eigenvalues converges towards the same deterministic distribution $\rho_T(b)$ and hence we can replace the linear statistics by integrals:
\begin{align}
\frac{1}{N}\sum_{i=2}^{N} f( \tilde{b}_i) &\approx  \int_{b_-}^{b_+}f(b) \rho_T(b) \mathrm{d}b  +o(1)  \, , &&
    \label{LinearStatIntegral}
\end{align}
which gives
\begin{align}
\begin{split}
\mathrm{Prob} \left[ b_1 <0 \right] \approx
     &      \int_{- \infty}^0 \mathrm{exp} \left[- \frac{N}{2} \left( \frac{1}{2T} \left( b_1 - b^{\star} \right)^2 -  \int_{b_-}^{b_+} \ln \sinh \left( \frac{\sigma}{2T} | b_1 - b| \right)   \rho_T(b) \mathrm{d}b  -  \int_{b_-}^{b_+} \ln | b_1 - b|   \rho_T(b) \mathrm{d}b \right) +o(N) \right] \mathrm{d}b_1    \\
     & \times \, \left( \int_{- \infty}^{0} \dots \int_{- \infty}^{0}  \mathcal{P}_{N-1}(b_2, \dots, b_N) \mathrm{d}b_2 \dots \mathrm{d}b_N \right) 
\end{split}
\label{eq:proba_stab.2}
\end{align}
The $(N-1)$-fold integral can be reduced to a constant by saddle-point approximation and in the end, we have:  
\begin{align}
\label{eq:Prob_int_LDP}
\mathrm{Prob} \left[ b_1 <0 \right] &\approx \int_{- \infty}^0 \mathrm{exp} \left[ - \frac{N}{2} \Psi_{\sigma,T}(b_1) +o(N) \right] \mathrm{d}b_1 \, ,&&
\end{align}
where, for every $w< b_{-}(T)$, the \emph{large deviation function} $ \Psi_{\sigma,T}(w) $ is given by: 
\begin{align}
\label{prop:LDF_wo_A}
 \Psi_{\sigma,T}(w)  &=  \frac{(w-b^{\star})^2}{2T} - \int_{b_-(T)}^{b_+(T)}  \ln(b-w)\rho_T(b) \mathrm{d}b  - \int_{b_-(T)}^{b_+(T)} \ln \sinh \left( \frac{\sigma}{2T}(b-w) \right)\rho_T(b)  \mathrm{d}b -A \, ,&&
\end{align}
with the constant $A$ chosen such that  $\Psi_{\sigma,T}(b_-(T))=0$, since  $b_1 {\to} b_-(T)$ as $ N \to \infty$. This gives:
\begin{align}
A &= \frac{(b_-(T)-b^{\star})^2}{2T} - \int_{b_-(T)}^{b_+(T)}  \ln( b- b_-(T))\rho_T(b) \mathrm{d}b - \int_{b_-(T)}^{b_+(T)}  \ln \sinh \left( \frac{\sigma}{2T}(b-b_-(T) ) \right)\rho_T(b) \mathrm{d}b \, . &&
\label{eq:constantofintegration}
\end{align}
Replacing the constant $A$ in Eq. \eqref{prop:LDF_wo_A} by its expression in Eq. \eqref{eq:constantofintegration}, one gets for the large deviation function: 
\begin{align}
\begin{split}
\label{prop:LDfunctionFlat}
 \Psi_{\sigma,T}(w) =& \,  \frac{(w-b^{\star})^2 - (b_-(T)-b^{\star})^2}{2T} \\&- \int_{b_-(T)}^{b_+(T)}  \ln \left(\frac{b-w}{b-b_-(T)} \right) \rho_T(b) \mathrm{d}b  - \int_{b_-(T)}^{b_+(T)}  \ln \left(\frac{\sinh ( \frac{\sigma}{2T}(b-w))}{\sinh  \frac{\sigma}{2T}(b-b_-(T))} \right) \rho_T(b) \mathrm{d}b \, . 
\end{split}
\end{align}
The function  $\Psi_{\sigma,T}(w)$ is decreasing on $(-\infty,0)$ and hence take its minimum at $0$, so that the integral in Eq. \eqref{eq:Prob_int_LDP} is dominated at large $N$ by the value at zero, which is nothing else than the left large deviation function we want to compute:
\begin{align}
  \Phi_{-}(\sigma,T) & = \Psi_{\sigma,T}(0)  \, . && 
  \label{eq:RateFunction}
\end{align}
Note that unlike the homogeneous case, corresponding to $\sigma=0$, one cannot simplify further this expression. Thus, we have 
\begin{align}
\mathrm{Prob} \left[ b_1 <0 \right] &\approx \mathrm{exp} \left[ - \frac{N}{2} \Phi_{-}(\sigma,T)  +o(N) \right]  \, ,&&
\label{eq:LDPfinal}
\end{align}
and from Eq. \eqref{eq:probstab.finalsec} the probability of stability writes:
\begin{align}
\label{eq:final_result}
 \mathcal{P}_{\mathrm{stable}}(T,N)&\approx 1 -\mathrm{exp} \left[ - \frac{N}{2} \Phi_{-}(\sigma,T) +o(N) \right] \, . &&
\end{align}
 This can be easily computed thanks to Eq. \eqref{eq:parametricsolution} for the density $\rho_T(b)$. A  plot of the large deviation function for $\sigma=1$ is given in Fig. \ref{fig:LDplot}.

\vskip 0.3 cm

\noindent \textit{Behavior of the rate function $\Psi_{\sigma,T}$ near the edge:} In this paragraph, we want to characterize the behavior of the function $\Psi_{\sigma,T}(w)$ near the edge $b_{-}(T)$, to see if one recovers the ‘$3/2$’ scaling as in Eq. \eqref{rate_asympt.1} so that it matches with the asymptotic behavior of the Tracy-Widom function given in Eq. \eqref{eq:TW_asymp}.   Let's consider  $w = b_{-}(T) - \epsilon$, with $\epsilon>0$ and $\epsilon << 1$. Looking at  small $\epsilon$ directly from Eq. \eqref{prop:LDfunctionFlat} is difficult, since one does not have an explicit expression for $\rho_T(b)$. It is always possible to integrate by part and do the change of variable $b_{\pm}(\rho) = b$ in Eq. \eqref{prop:LDfunctionFlat} to make the integrals depend only on the parametric solution  $b_{\pm}(\rho)$, for which one has an explicit solution \eqref{eq:parametricsolution}. However, this will give  an involved expression, and it is hard to get the asymptotic at small $\epsilon$ from it. Instead, the idea is to write the large deviation function as an integral of length $\epsilon$ so that at first order in $\epsilon$ we can discretize it with Euler's method. Let's first notice the following integral representation for the rate function,
\begin{align}
\label{eq:Psi_integral}
\Psi_{\sigma,T}(w) &= - \int_{w}^{b_{-}(T)} ( \Psi_{\sigma,T})'(s) \, \mathrm{d}s \, , &&
\end{align}
where the constant of integration is zero since $\Psi_{\sigma,T}(b_{-}(T))=0$. From Eq. \eqref{prop:LDF_wo_A}, the derivative of the rate function can be written as: 
\begin{align}
    ( \Psi_{\sigma,T})'(s) &= \frac{\mathrm{d}}{\mathrm{d}s} \left[ \frac{(s-b^{\star})^2}{2T} - \int_{b_-(T)}^{b_+(T)}  \ln(b-s)\rho_T(b) \mathrm{d}b  - \int_{b_-(T)}^{b_+(T)} \ln \sinh \left( \frac{\sigma}{2T}(b-s) \right)\rho_T(b)  \mathrm{d}b  \right] \, , &&  \label{eq:dPsi.1} \\
     ( \Psi_{\sigma,T})'(s) &= \frac{\mathrm{d}}{\mathrm{d}s} \left[ \frac{(s-b^{\star})^2}{2T} - \int_{b_-(T)}^{b_+(T)}  \ln(b-s)\rho_T(b) \mathrm{d}b  - \int_{b_-(T)}^{b_+(T)} \ln  \left( \mathrm{e}^{\frac{\sigma}{T}b}- \mathrm{e}^{\frac{\sigma}{T}s} \right)\rho_T(b)  \mathrm{d}b + \frac{\sigma}{T}s \right] \, , &&  \label{eq:dPsi.2}
\end{align}
where to go from Eq. \eqref{eq:dPsi.1}  to Eq. \eqref{eq:dPsi.2}, we have used once again the  trigonometric identity \eqref{iden_sinh} and discarded terms which do not depend on the variable $s$. Since $b^{\star}=1 + \frac{\sigma}{2}$, this can be equivalently written as: 
\begin{align}
    ( \Psi_{\sigma,T})'(s) &= \frac{\mathrm{d}}{\mathrm{d}s} \left[ \frac{(s-1)^2}{2T} - \int_{b_-(T)}^{b_+(T)}  \ln(b-s)\rho_T(b) \mathrm{d}b   - \int_{b_-(T)}^{b_+(T)} \ln  \left( \mathrm{e}^{\frac{\sigma}{T}b}- \mathrm{e}^{\frac{\sigma}{T}s} \right)\rho_T(b)  \mathrm{d}b \right] \, , &&  \label{eq:dPsi.3} \\
      ( \Psi_{\sigma,T})'(s) &= \frac{1}{T}(s-1) - g(s,T) - \frac{\sigma}{T} \mathrm{e}^{\frac{\sigma}{T}s} \int_{b_-(T)}^{b_+(T)} \frac{\rho_T(b) \mathrm{d}b}{\mathrm{e}^{\frac{\sigma}{T}s} - \mathrm{e}^{\frac{\sigma}{T}b}} \, . &&  \label{eq:dPsi.4}
\end{align}
To simplify the last term in Eq. \eqref{eq:dPsi.4}, let's introduce the matrix $\mat{C} = \mathrm{exp} \left\{ \frac{\sigma}{T} \mat{B} \right\} $. The average density of the eigenvalues $\{ c_i \} $ of the matrix $\mathbf{C}$
\begin{align}
   \nu(c,N)&= \frac{1}{N}\, \left\langle \sum_{i=1}^N \delta(c_i-c)\right\rangle \, , &&
\end{align}
is related, in the large $N$ limit, to the spectral density $\rho_T$ by:
\begin{align}
    \nu(c) &=  \nu(c,N \to \infty) = \frac{T}{\sigma} \frac{\rho_T \left( \frac{T}{\sigma} \ln c \right)}{c} \, . &&
\label{eq:law_exp_B}
\end{align}
Similarly, the edges $c_{\pm}$ of the density $\nu(c)$ are given in terms of the edges $b_{\pm}(T)$ of the matrix $\mathbf{B}$ by:
\begin{align}
    c_{\pm} &=e^{ \frac{\sigma}{T} b_{\pm}(T)} \,. && 
    \label{eq:edges_C}
\end{align}
If we introduce the resolvent $g_C$ of the matrix $\mathbf{C}$:
\begin{align}
    g_C(s) &= \int_{\mathrm{c}_{-}}^{\mathrm{c}_{+}} \frac{\nu(c) \mathrm{d}c}{s-c}  \, ,&&
    \label{def:resolvent_of_C}
\end{align}
one can rewrite Eq. \eqref{eq:dPsi.4} as:
\begin{align}
      ( \Psi_{\sigma,T})'(s) &= \frac{(s-1)}{T} - g(s,T) - \frac{\sigma}{T} \mathrm{e}^{\frac{\sigma}{T}s} g_C(\mathrm{e}^{\frac{\sigma}{T}s}) \, . &&  \label{eq:dPsi.5}
\end{align}
Using the integral representation Eq. \eqref{eq:Psi_integral} for $w = b_{-}(T) - \epsilon$ together with Eq. \eqref{eq:dPsi.5}, we can write the large deviation as:
\begin{align}
    \Psi_{\sigma,T}(b_-(T) -\epsilon) &= - \int_{b_{-}(T) - \epsilon }^{b_{-}(T)}  \left( \frac{(s-1)}{T} -g(s,T) - \frac{\sigma}{T} \mathrm{e}^{\frac{\sigma}{T}s} g_C(\mathrm{e}^{\frac{\sigma}{T} s}) \right) \mathrm{d}s  \, .   && 
    \label{eq:Euler.1}
\end{align}
 We can now approximate the integral by Euler's methods: 
\begin{align}
\Psi_{\sigma,T}(b_-(T) -\epsilon) &\underset{\epsilon \to 0^+}{\sim} \epsilon \, \times \left( \frac{( 1 - b_-(T) + \epsilon)}{T} + g( b_-(T) - \epsilon,T) + \frac{\sigma}{T} \mathrm{e}^{\frac{\sigma}{T}( b_-(T) - \epsilon)} g_C(\mathrm{e}^{\frac{\sigma}{T} ( b_-(T) - \epsilon)}) \right) +O(\epsilon^2) \, . \label{eq:Euler.2}&&
\end{align}
Next, we need the behavior of the resolvents close to the edges: 
\begin{itemize}
    \item since the edge $b_-(T)$ is solution of Eq. \eqref{prop:edge_from_resolvent}, the inverse function $z(\mathrm{g})$ is locally quadratic around the point $g(b_{-}(T))$, see Fig. \ref{fig:inverse_resolvent}: 
\begin{align}
    z(\mathrm{g}) &\sim b_-(T) + C \mathrm{g}^2 \quad \mbox{ for } \mathrm{g} \to g(b_{-}(T)) \, , &&
    \label{eq:behavior_zg_edge}
\end{align}
where $C$ is a constant. Inverting this relation, the resolvent behaves as a square root near the edge: 
\begin{align}
    g( b_-(T) - \epsilon,T) &\sim g( b_-(T),T) + \alpha_B \sqrt{\epsilon} \quad \mbox{ for } \epsilon \to 0^+ \, , &&
    \label{eq:behavior_res_edge}
\end{align}
where $\alpha_B$ is a constant of proportionality.
\item  Similarly, since $c_-  =\mathrm{e}^{\frac{\sigma}{T} b_-(T)}$ is the edge  of the distribution $\nu(c)$ so we have: 
\begin{align}
    g_C \left(\mathrm{e}^{\frac{\sigma}{T} ( b_-(T) - \epsilon)} \right) &\sim g_C( c_- ) + \alpha_C \sqrt{\epsilon} \quad \mbox{ for } \epsilon \to 0^+ \, , &&
    \label{eq:behavior_res_edge_C}
\end{align}
where $\alpha_C$ is a constant of proportionality.
\end{itemize}
This gives for the large deviation function: 
\begin{align}
\Psi_{\sigma,T}(b_-(T) -\epsilon) &\sim \left( \frac{( 1- b_-(T))}{T}  + g(  b_-(T),T ) + \frac{\sigma}{T}  c_{-} g_C \left(c_{-} \right) \right) \epsilon + \alpha \epsilon^{\frac{3}{2}} +O(\epsilon^2) \quad \mbox{ for } \epsilon \to 0^+  \, , &&
\label{eq:Euler.3}
\end{align}
but using once again the trigonometric relation \eqref{iden_sinh} together with the expression for the bottom edge Eq. \eqref{eq:edges_C}, the linear term can be expressed as: 
\begin{align}
     \frac{( 1- b_-(T))}{T}  + g(  b_-(T),T ) + \frac{\sigma}{T}  c_{-} g_C \left(c_{-} \right) &= \frac{( b^{\star}- b_-(T))}{T}  + \int \frac{\rho_T(b)}{b_{-}(T) -b} \mathrm{d}b + \frac{\sigma}{2T} \int \rho_T(b) \coth \left( b_{-}(T) -b \right) \mathrm{d}b \,  .&&
\end{align}
This exactly the solution of the saddle-point equation \eqref{eq:SP_density} for the edge $b_-(T)$. As a consequence, this term is exactly zero. Thus, we get the correct ‘$3/2$’ scaling, as expected: 
\begin{align}
    \Psi_{\sigma,T}(b_-(T) -\epsilon) & \propto \epsilon^{\frac{3}{2}} \quad \mbox{ for } \epsilon \to 0^+  \, . &&
    \label{eq:BehaviorRateFunction_edge}
\end{align}

 \section{Conclusion}
 \label{sec:Conclusion}

In this paper, we have studied the probability of stability of 
a large complex system of size $N$ within the framework of a 
generalized May model, which takes into account a possible 
heterogeneity $a_i \neq a_j$ in the intrinsic relaxation rates of each species. 
In this model, the control parameter is $T$ which is the square
of the interaction strength of the random pairwise interaction
between the different species.
For generic distribution $\mu(a)$ of the $a_i$'s, Eq. 
\eqref{prop:equationTc} completely characterizes the critical point 
$T_c$ of the May-Wigner phase transition, where
the system undergoes a transition from a `stable' phase
to an `unstable' phase as $T$ increases. Focusing on the special case 
where the $a_i$'s follow what we call the flat initial condition 
\eqref{flatIC}, where $\sigma$ is the only new parameter of the model 
controlling the spread of the distribution $\mu(a)$, we are able to 
(i) characterize how $T_c$ behaves with $\sigma$, (ii) to obtain the parametric 
solution of the eigenvalue density of stability matrix in the large $N$ 
limit for any $T$, and (iii) to obtain the `left' large deviation function
$\Phi_{-}(\sigma,T)$ that controls 
the probability of stability for  $T<T_c$ on the stable side, for
large but finite $N$. One important 
challenge is to develop a framework to compute the `right' large deviation 
function $\Phi_+(\sigma,T)$ which characterizes the probability of 
stability in the 
unstable phase $(T> T_c)$. To compute $\Phi_+$,
one needs to find the equilibrium 
measure of a \emph{pushed-to-the-origin} gas of particles with a mixture of 
logarithmic and log-sinh pairwise interactions, as given in the 
joint law of eigenvalues. This remains out of reach.
Finally, another natural question is to 
investigate the large 
deviation function for other initial conditions, for which we do not 
have a simple formula for the joint law of eigenvalues.

\section*{Acknowledgments}

We thank Tristan Gautié, Pierre le Doussal, Marc Potters and
Gregory Schehr for useful discussions.

\bibliographystyle{ieeetr}
\bibliography{biblio}


\appendix

\section{Properties of the resolvent}
\label{sec:Ap:prop_res}
In this section, we recall the main properties of the resolvent.
\subsection{definition}
\label{sec:Ap:def_res}
 For a probability distribution $\rho$, its resolvent - also known as the \emph{Green function} or \emph{Stieltjes transform} - is defined as: 
\begin{align}
\label{def:Ap:res}
    g_{\rho}(z) &= \int \frac{\rho(x)}{z-x} \mathrm{d}x  \, . &&
\end{align}
The resolvent is defined for every $z$ in  the complex plane except for values $z=x$ on the real line such that $\rho(x)>0$, otherwise the integral diverges.
\begin{itemize}
    \item In particular, if $\rho$ is a smooth density function supported on an interval $[x_-,x_+]$, where $x_{\pm}$ is the top/bottom edge, then its resolvent is defined for all $z \in \mathbb{C} \setminus [x_-, x_+]$
    \begin{align}
    \label{def:Ap:res_smooth}
        g_{\rho}(z) &= \int_{x_-}^{x_+} \frac{\rho(x)}{z-x} \mathrm{d}x \, .&&
    \end{align}
    \item  Another particular case is the case where $\rho$ is a discrete measure of the form $\rho(x) = \frac{1}{N} \sum_{i=1}^N \delta(x - x_i)$ which appear naturally in the context of random matrices with $\vect{x}=(x_1, \dots, x_N)$ the vector of eigenvalues of a matrix. In this case, the resolvent is defined for all $z \in \mathbb{C} \setminus \left\{ \vect{x} \right\}$ and be simply written as:
     \begin{align}
     \label{eq:Ap:res_discrete}
        g_{\rho}(z) &= \frac{1}{N} \sum_{i=1}^N \frac{1}{z-x_i}  \, . &&
    \end{align}
\end{itemize}
\noindent \emph{Resolvent of the semi-circular law:} The Wigner semi-circular probability density with variance one is the average density of eigenvalues of GOE matrices as $N \to \infty$. For all $x$ in $[-2,2]$, this density is given by: 
\begin{align}
\label{eq:Ap:wignerdensity}
    \rho_{\mathrm{Wig}}(x) &= \frac{1}{2 \pi} \sqrt{4 - x^2} \, .&&
\end{align}
From Eq. \eqref{def:Ap:res} and integrating, one gets for the resolvent if the Wigner semi-circular distribution: 
\begin{align}
\label{res_sc}
    g_{\mathrm{Wig}}(z) &= \frac{1}{2} \left( z- z\sqrt{1-\frac{4}{z^2}} \right) \, .&&
\end{align} 

\subsection{Inversion formula}
\label{sec:Ap:InversionFormula}
From a resolvent, one can get back the distribution by looking at the imaginary part of the resolvent close to the support of the distribution. We have: 
\begin{align}
\label{eq:Ap:ImResolvent}
    \mathfrak{Im} g( x - \mathrm{i} \epsilon) &= \int \rho(x') \mathfrak{Im} \frac{1}{x-\mathrm{i}\epsilon -x'} \mathrm{d}x' \, . &&
\end{align}
Multiplying the fraction by the conjugate of the denominator and taking only the imaginary part, one gets: 
\begin{align}
\label{eq:Ap:ImResolvent.2}
     \mathfrak{Im} g( x - \mathrm{i} \epsilon) &=  \pi \left( \rho \ast  P_{\epsilon} \right)(x) \, . && 
\end{align}
The symbol $\ast$ denotes the classical convolution product : $(f \ast g) (x) = \int f(x') g(x-x') \mathrm{d}x'$ and $P_{\epsilon}$ is the \emph{Cauchy kernel} defined by
\begin{align}
\label{eq:Ap:CauchyKernel}
    P_{\epsilon}(x) &= \frac{1}{\pi} \frac{\epsilon}{x^2 + \epsilon^2} \, .&& 
\end{align}
 The kernel $P_{\epsilon}$ is the probability density of a centered Cauchy random variable with width $\epsilon$. As the width $\epsilon \to 0^+$, we have
\begin{align}
\label{eq:Ap:CauchytoDirac}
    P_{\epsilon}(x) &\underset{\epsilon \to 0^+}{\to} \delta(x) \, , &&
\end{align}
where $\delta$ is a Dirac mass function. Thus, if one combines Eq. \eqref{eq:Ap:ImResolvent.2} and Eq. \eqref{eq:Ap:CauchytoDirac}, one gets the Socochi-Plemelj inversion formula: 
\begin{align}
\label{eq:Ap:SPinvformula}
    \rho(x) &= \frac{1}{\pi}  \mathfrak{Im} \, g( x - \mathrm{i} 0^+)  \, .&&
\end{align}

\subsection{edges and resolvent}
\label{Sec:App:edge_and_res}
\begin{figure}
     \centering
     \begin{subfigure}[b]{0.49\textwidth}
         \centering
         \includegraphics[width=\textwidth]{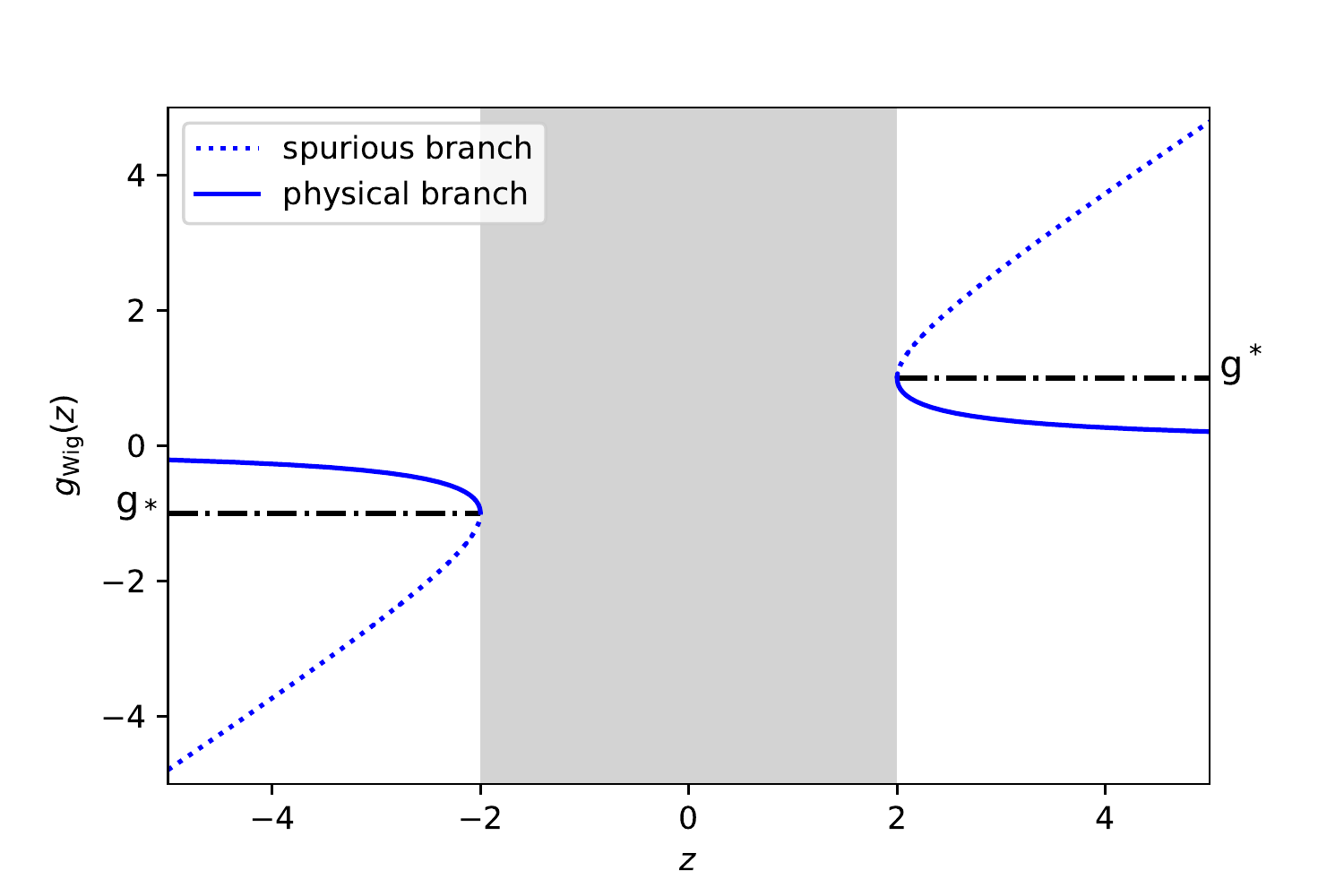}
     \end{subfigure}
     \hfill
     \begin{subfigure}[b]{0.49\textwidth}
         \centering
         \includegraphics[width=\textwidth]{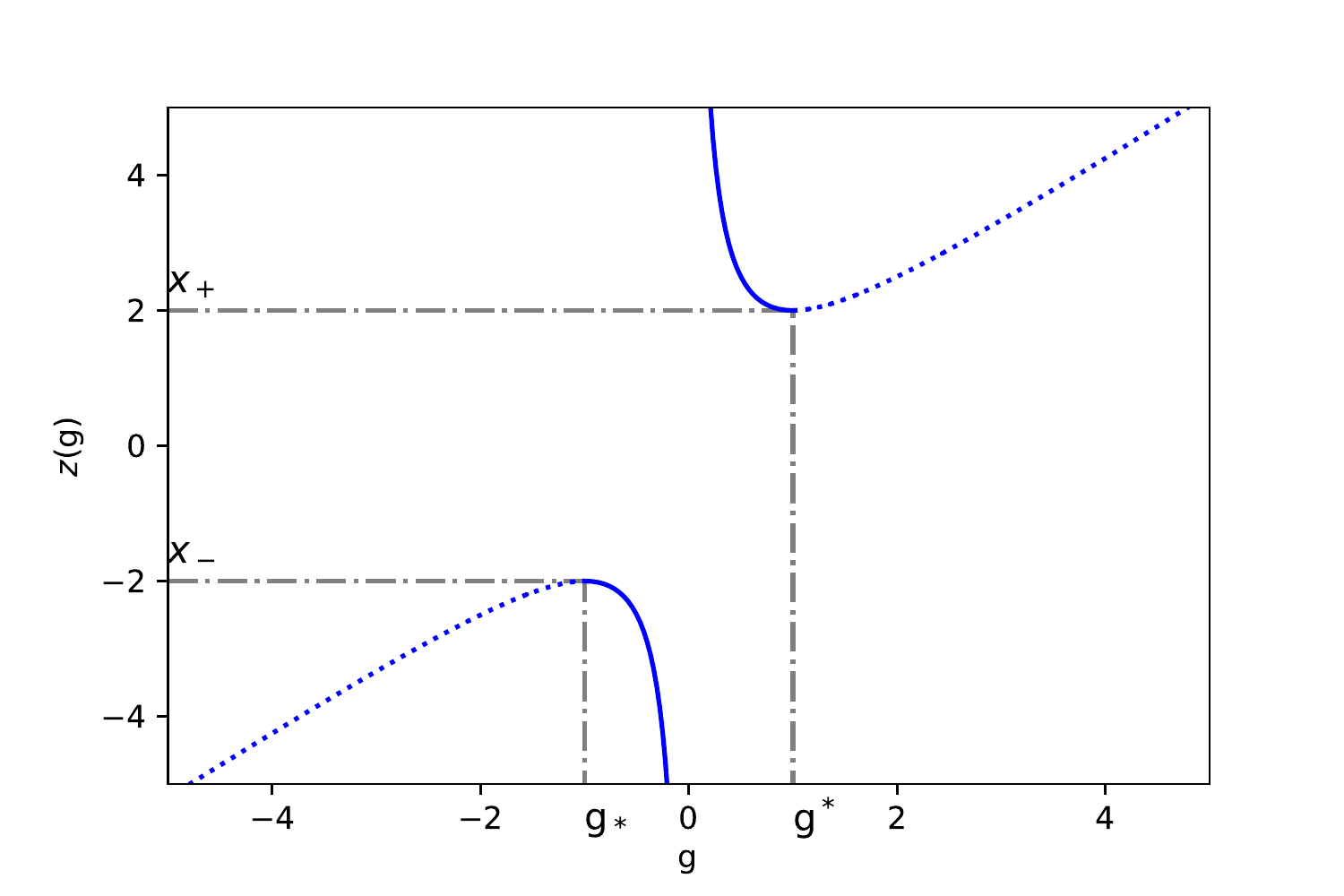}
     \end{subfigure}
     \caption{ \textbf{(Left)} Plot of the resolvent for the semi-circular distribution for different value of $z$. The gray rectangle represents the region where the resolvent is not defined on the real line. The solid-line curve is the correct representation of the resolvent as given in Eq. \eqref{res_sc} while the dotted line  is the non-physical branch obtained by inverting the  function $z(\mathrm{g}) = \mathrm{g}+ \frac{1}{\mathrm{g}}$ for values of $\mathrm{g}$ outside $[-1,1]$. \textbf{(Right)} Plot of the inverse function $z(\mathrm{g})=\mathrm{g}+\frac{1}{\mathrm{g}}$ of the resolvent of the semi-circular distribution. The dotted-line curve represents the analytical extension of the function outside the interval $[-1,1]$.}
     \label{fig:inv_stieltjes}
\end{figure}

Let's consider that one has an unknown smooth density probability $\rho$, and the goal is to compute the edges $x_{\pm}$ of this density from the knowledge of the resolvent. Fix $z$ on the real line outside the support $[x_-, x_+]$ where the resolvent is ill-defined. Let's first notice that if we derive Eq. \eqref{def:Ap:res_smooth} with respect to $z \in \mathbb{R}$, we get: \begin{align}
    g'(z) &= - \int_{x_-}^{x_+} \frac{\rho(x)}{(z-x)^2} <0 \, , &&
\end{align} 
and therefore the resolvent is strictly decreasing in each region $(- \infty,x_-)$ and $(x_+,\infty)$, and from Eq. \eqref{def:Ap:res_smooth}. Furthermore, we have $\lim_{|z|\to \infty} g(z) =0$. Since it is decreasing on each interval, for a fixed value $\mathrm{g}$, we can always find an inverse function $z(\mathrm{g})$, also decreasing, such that $g(z(\mathrm{g})) =\mathrm{g}$.  
\vskip 0.3cm
\noindent \textit{Inverse resolvent of the semi-circular law:} as a concrete example, let's look at the semi-circular  distribution whose resolvent is given in Eq. \eqref{res_sc}. For $z$ real and outside the support $[x_-, x_+]$, finding the inverse function $z(\mathrm{g})$ corresponds to invert the relation \eqref{res_sc} and one gets: 
\begin{align}
\label{inv_res_sc}
    z(\mathrm{g}) &= \mathrm{g} + \frac{1}{\mathrm{g}}  \, .&&
\end{align}
\vskip 0.3cm
\noindent The function $z(\mathrm{g})$ is theoretically only defined between\footnote{except at $0$ since we have  $\lim_{|z|\to \infty} g(z) =0$ and hence $z(\mathrm{g})$ diverges at $0$.} the two values $\mathrm{g}_*$ and $\mathrm{g}^*$ such that $z(\mathrm{g}_*) = x_-$ and $z(\mathrm{g}^*) = x_+$. Now, it turns out that one can generally extend this function $z(\mathrm{g})$ for values outside this region. The inverse resolvent of the semi-circular law is a clear example of such a case, since Eq. \eqref{inv_res_sc} makes sense for any $\mathrm{g} \neq 0$. However, just after the point $\mathrm{g}^*$ (or $\mathrm{g}_*$), $z(\mathrm{g})$ can not continue to be monotonic: otherwise one can invert it again to get back the resolvent $g(z)$ which would be a real function in the support of the probability distribution. This in contradiction with the Socochi-Plemelj formula \eqref{eq:Ap:SPinvformula}, see Fig. \ref{fig:inv_stieltjes}. Hence, the derivative of the function $z(\mathrm{g})$ must cancel at the point $\mathrm{g}^*$ (resp. $\mathrm{g}_*$). Since this is the point where it takes the value $x_+$ (resp. $x_-$), we have the following properties for the edges: 
\begin{align}
\label{eq:Ap:edgeinvres1}
   x_{-} = z(\mathrm{g}_*) \quad &\text{and} \quad  x_{+} = z(\mathrm{g}^*) \, ,&&
\end{align}
where $\mathrm{g}_*$ and $\mathrm{g}^*$ are respectively the lowest and highest solution of: 
\begin{align}
\label{eq:Ap:edgeinvres2}
    z'(\mathrm{g})&=0 \, . &&
\end{align}

\noindent \textit{Recovering the edges of  the semi-circular distribution:} Using  Eq. \eqref{eq:Ap:edgeinvres2} in the case of the semi-circular distribution where the inverse resolvent is given in Eq. \eqref{inv_res_sc}, one gets $\mathrm{g}^* = 1 = - \mathrm{g}_*$ and hence by Eq. \eqref{eq:Ap:edgeinvres1}, the edges are  $x_{\pm} = \pm 2$, as expected. 

\vskip 0.5cm

\end{document}